\documentclass[twocolumn, dvipsnames]{aastex631}
\usepackage{amsmath}
\usepackage{graphicx}
\usepackage{graphbox}
\usepackage{tikz}
\usepackage{float}
\usepackage{multirow}
\usepackage{color}
\usepackage{makecell}
\usepackage{wrapfig}
\usepackage{comment}

\accepted{in AJ - March 2024}

\shorttitle{Panchromatic light-curve retrievals: \textsc{ExPLOR}.}
\shortauthors{Changeat et al. 2024}

\begin{document}
	\title{Towards atmospheric retrievals of panchromatic light-curves: \textsc{ExPLOR}-ing generalized inversion techniques for transiting exoplanets with JWST and Ariel.}
	\correspondingauthor{Q. Changeat}
	\email{qchangeat@stsci.edu}
	\author[0000-0001-6516-4493]{Q. Changeat $^\dagger$}
        \affil{European Space Agency (ESA), ESA Office, 
        Space Telescope Science Institute (STScI), Baltimore MD 21218, USA.}
	    \affil{Department of Physics and Astronomy,
		University College London,
		Gower Street,WC1E 6BT London, United Kingdom}
  
    \author[0000-0002-0598-3021]{Y. Ito}
    \affil{National Astronomical Observatory of Japan, Osawa 2-21-1, Mitaka, Tokyo 181-8588, Japan}

    \author[0000-0003-2241-5330]{A. F. Al-Refaie}
    \affil{Department of Physics and Astronomy,
		University College London,
		Gower Street,WC1E 6BT London, United Kingdom}

    \author[0000-0002-9616-1524]{K. H. Yip}
    \affil{Department of Physics and Astronomy,
		University College London,
		Gower Street,WC1E 6BT London, United Kingdom}

    \author[0000-0000-0000-0001]{T. Lueftinger}
    \affil{European Space Agency (ESA), European Space Research and Technology Centre (ESTEC), 2201 AZ Noordwijk, The Netherlands}

\renewcommand{\thefootnote}{\fnsymbol{footnote}}
\footnotetext[2]{ESA Research Fellow}
\renewcommand{\thefootnote}{\arabic{footnote}}

\begin{abstract}
    Conventional atmospheric retrieval codes are designed to extract information, such as chemical abundances, thermal structures and cloud properties, from fully ``reduced'' spectra obtained during transit or eclipse. Reduced spectra, however, are assembled by fitting a series of simplified light-curves to time series observations, wavelength-by-wavelength. Thus, spectra are a post-processed summary statistics of the original data, which by definition does not encode all the available information (i.e., astrophysical signal, model covariance, instrumental noise). Here, we explore an alternative inversion strategy where the atmospheric retrieval is performed on the light-curve directly -- i.e., closer to the data. This method is implemented in \textsc{ExPLOR} (EXoplanet Panchromatic Light-curve Observation and Retrieval), a novel atmospheric retrieval code inheriting from the \textsc{TauREx} project. By explicitly considering {\it time} in the model, \textsc{ExPLOR} naturally handles transit, eclipse, phase-curve and other complex geometries for transiting exoplanets. In this paper, we have validated this new technique by inverting simulated panchromatic light-curves. The model was tested on realistic simulations of a WASP-43\,b like exoplanet as observed with the James Webb Space Telescope (JWST) and Ariel telescope. By comparing our panchromatic light-curve approach against conventional spectral retrievals on mock scenarios, we have identified key breaking points in information and noise propagation when employing past literature techniques. Throughout the paper, we discuss the importance of developing ``closer-to-data'' approaches such as the method presented in this work, and highlight the inevitable increase in model complexity and computing requirements associated with the recent JWST revolution.
\end{abstract}

\keywords{Exoplanet atmospheres (487); Astronomy data analysis (1858); Time series analysis(1916); James Webb Space Telescope (2291)}


\section{Introduction}

\underline{\it Context:} In the field of exoplanetary atmospheres, the so-called {\it atmospheric retrieval} technique stands out as a paramount approach for extracting the information from low to medium resolution (i.e., R $<$ 5,000) spectroscopic observations. Originally conceived to address the lack of prior knowledge and fully exploit the limited capabilities of past telescopes, this inversion technique is now considered as an unbiased method to explore the extensive and highly degenerate parameter space of exoplanet models \citep{Madhu_retrieval_method}. Notably, improvements in the theory \cite[e.g.,][]{Line_2013, Line_2014_syst, Waldmann_taurex2, Waldmann_taurex1, Madhu_2018_ret} as well as the standardization of the core algorithms -- facilitated by the availability of open-source software \citep{Waldmann_taurex1,  Lavie_2017, Villanueva_2018, 2019_al-refaie_taurex3, Molliere_2019, Zhang_2019, Harrington_2022} -- have permitted the establishment of those sophisticated and robust tools across the exoplanet community. In parallel, benchmarking initiatives \citep{Barstow_2020} have played a crucial role in establishing some level of consensus for at least the simplest features. Nowadays, most {\it quantitative} observational studies of exo-atmospheres incorporate insights from atmospheric retrievals in one form or another \cite[see e.g., population studies and recent JWST works: ][]{tsiaras_30planets, fisher, pinhas, Welbanks_2019, Min_2020, Cubillos_2021_pop, Roudier_2021, Changeat_2022_five, edwards_pop, Estrela_2022, August_2023, Coulombe_2023, Jiang_2023, Kempton2023, Moran_2023, Taylor_2023, Edwards_2024, Dyrek_2024}. 

Many teams have engaged in developing exoplanet retrieval codes \citep{Rengel_2023}. Despite those diverse efforts, there is a general trend towards standardizing the technique. Most of the codes are one-dimensional (1D), typically modeling a particular atmospheric region (e.g., the dayside or the terminator chord), and focusing on the inversion of eclipse or transit observations. However, most atmospheric studies today involve tidally locked exoplanets, which have thermally and chemically in-homogeneous atmospheres driven by the strong irradiation contrast experienced by their day and night sides -- a phenomenon confirmed in theoretical studies \citep{Showman_Guillot_2002, Cho_2003, Skinner_2022_cyclogenesis}. To properly characterize those processes, more constraining phase-resolved observations are typically necessary. Spectroscopic phase-curve, however, requires significant telescope time, instrument stability, and involve more intricate analyses surpassing the capabilities of 1D models. Such difficulties could explain why only a handful of targets have been observed by the Hubble Space Telescope (HST). So far, HST has only observed the phase-curves of hot Jupiters, such as WASP-43\,b \citep{Stevenson_2014}, WASP-103\,b \citep{Kreidberg_w103}, WASP-18\,b \citep{arcangeli_w18_phase}, and WASP-121\,b \citep{Evans_2022_diunarl}. In the near future, the excellent pointing stability and moderate instrument systematics of JWST \citep{Rigby_2023, Espinoza_2023} will allow to fully exploit the richness of spectroscopic phase-curves, which are now gaining popularity\footnote{The following phase-curves are already planned with JWST: WASP-43\,b, WASP-121\,b, GJ-1214\,b, K2-141\,b, NGTS-10\,b, GJ-367\,b, LTT-9779\,b, TOI-561\,b, TOI-1685\,b, K2-22\,b, and TOI-2445\,b}

In the HST era, initial approaches to interpret phase-resolved spectroscopic data consisted in performing 1D emission retrievals at discretized phases \citep{Stevenson_2014}. However, this method falls short of maximizing the information content of those datasets and is susceptible to 3D biases \citep{Changeat_2020_phasecurve1}. Recently, more sophisticated methodologies have emerged. For instance, \cite{irwin_2020_w43, Yang_2023} presented some of the first unified methods dedicated to spectroscopic phase-curve data. They fitted all the observed phases of the WASP-43\,b HST data \citep{Stevenson_2014} using a single 2.5D model with optimal estimation. \cite{Cubillos_2021} developed a method to extract longitudinally resolved spectra, allowing more accurate 1D retrievals to be performed. Other works have successfully performed three-dimensional atmospheric retrievals on phase-curve data \citep{Chubb_2022}, highlighting the computational challenges arising from the augmented information content (e.g., transmission spectra vs panchromatic phase-curves) and the need for more physically motivated prescriptions to manage the increased problem dimension. Finally, some works have adopted simplified geometry assumptions, for instance by dividing the planet into distinct atmospheric regions \citep{Changeat_2020_phasecurve1, Feng_2020_2D, Evans_2022_diunarl}, to accelerate the forward modeling and reduce the overall computational requirement. Despite their novelty, all those techniques act on transformed representations of the data (i.e., {\it spectra}). In this work, we discuss the advantages of performing atmospheric retrievals closer to the data, presenting a new code tailored to invert atmospheric information directly from panchromatic light-curves of exoplanet observations. \\

\underline{\it Motivation:} {\it why retrieving at the light-curve level?} Following the correction of the detector images and the processing of the ramps, the current approach is to perform a pre-retrieval reduction stage. This stage involves fitting some simplified light-curve model convolved with an instrument systematic model. The goal is to remove the remaining instrument systematics and simplify the subsequent atmospheric retrieval step by eliminating the time component from the data. Despite its apparent advantage, this two step process -- i.e., performing a light-curve reduction before conducting atmospheric retrievals -- can introduce many issues and biases, a few of which are noted below: 

\begin{description}\itemsep0em 
    
\item[{\small Wavelength binning}] wavelength binning at the light-curve stage is often required to boost the transit/eclipse signature and anchor the light-curve fit. Such binning dilutes the information contained in individual wavelength channels and leads to an overall loss of information content.

\item[{\small De-correlated fits}] typical reduction procedures are performed using wavelength-by-wavelength light-curve fits. Since each fit uses a completely independent Monte Carlo or Nested Sampling parameter exploration routine, this ignores the correlation between wavelength channels probing common processes. For instance if a molecule has spectral features in two separated wavelength channels, a simultaneous fit of both channels would allow to extract more information than performing two separated, independent, fits. As such, wavelength-by-wavelength light-curve fits leads to information loss.

\item[{\small Noise properties}] Spectra are often constructed from the mean and variance of fitted parameters (e.g., transit depth, or eclipse depth) rather than their full distribution. Using those ``summary statistics'' as observables for retrievals is incorrect from a noise propagation perspective.

\item[{\small Parameter Conditioning}] In current strategies, orbital and instrument systematics parameters are fitted separately (at the reduction stage) from the bulk planet and atmospheric parameters (at the retrieval stage). This leads to parameter conditioning and could bias our interpretations.

\item[{\small Complex observations}] Repeated observations of potentially variable atmospheres, phase-curve observations, eclipse mapping, and other complex events (e.g., multi-planet transit, planet+moon systems) cannot be naturally modeled using atmospheric retrievals codes that do not include {\it time}.

\item[{\small Contamination}] more generally, time dependent astrophysical signals such as stellar activity, or spot crossing event are difficult to disentangle when they are not simultaneously modeled with the planetary atmosphere.

\end{description}

Despite the obvious need for a more fundamental and comprehensive analysis techniques of exo-atmospheric observations, no atmospheric retrieval strategy expect \cite{Yip_2020_LC} has attempted to explicitly include time ($t$). Including time in atmospheric retrievals provides a natural solution to the above issues. \cite{Yip_2020_LC} pioneered atmospheric light-curve retrievals by demonstrating their relevance for transits. Specifically, they explored the impact of instrument systematics and orbital elements mismatch when fitting light-curve data from different telescopes, showing the precariousness of spectral retrievals combining datasets. \\

\underline{\it This work:} in this study, we generalize the work from \cite{Yip_2020_LC} and introduce a novel inversion approach tailored for interpreting spectroscopic observations of transiting exoplanets at the light-curve level: the {\it panchromatic light-curve retrieval}. Specifically, we leverage the \textsc{TauREx3.1} atmospheric phase-curve model from \cite{changeat_2021_phasecurve2}, which was designed to efficiently handle phase-dependent emission without undue computational scaling with the number of phases. This capability is integrated with the transit/eclipse light-curve models of \textsc{PyLightcurve}, resulting in a new atmospheric retrieval model \textsc{ExPLOR} ({\it EXoplanet Panchromatic Light-curve Observation and Retrieval}). \textsc{ExPLOR} introduces time as a fundamental component of the atmospheric retrieval, allowing us for the first time to generalize exo-atmospheric fits to spectro-temporal flux data. In Section \ref{sec:meth}, we provide an overview of the code \textsc{ExPLOR} and outline the methodology employed in this study. In Section \ref{sec:res}, we show examples of panchromatic light-curve retrievals with \textsc{ExPLOR}, with a focus on a WASP-43\,b-like exoplanet. Section \ref{sec:disc} discusses the advantages of our approach, emphasizing on a scenario where the information is diluted by the conventional reduction + retrieval strategy. Finally, in Section \ref{sec:conc}, we draw the conclusions of this work and highlight some potential next steps.

\section{Software and Methodology} \label{sec:meth}

\subsection{Model}
For this project, we have developed a new model \textsc{ExPLOR} to invert exoplanet atmospheric properties from panchromatic light-curve data. The code employs a similar class structure to \textsc{TauREx3.1}\footnote{\textsc{TauREx3}: \url{https://github.com/ucl-exoplanets/TauREx3_public}} \citep{2019_al-refaie_taurex3, al-refaie_2021_taurex3.1}. It combines the recently developed 1.5D phase-curve plugin \citep{Changeat_2020_phasecurve1, changeat_2021_phasecurve2, changeat_2021_w103} with the open-source light-curve package \textsc{PyLightcurve}\footnote{\textsc{PyLightcurve}: \\ \url{https://github.com/ucl-exoplanets/pylightcurve}} \citep{tsiaras_plc}. By combining the capabilities of those two libraries, \textsc{ExPLOR} considers time ($t$) as a new component in the atmospheric model, enabling full Bayesian retrieval of panchromatic light-curve observations (including events such as limb-darkened transits and eclipses). The role of each components of the code is described below:\\

\underline{\it Modeling the atmospheric flux along the orbit}: The planetary emission is modeled along $t$ using the \textsc{TauREx3} 1.5D phase-curve model \citep{changeat_2021_phasecurve2}. This model simulates an atmosphere as three separated regions (hotspot, dayside and nightside) of homogeneous properties and computes the emission contribution of each region at each phase using a quadrature integration scheme. Interested readers can refer to \cite{Changeat_2020_phasecurve1, changeat_2021_phasecurve2, changeat_2021_w103}, but we here summarize the main features of the model. As described in the aforementioned works, each region has distinctive properties -- thermal structure, chemical profiles, cloud properties -- which can be set separately or coupled in a flexible manner. While relatively simple compared to the complex 3D structures of exo-atmospheres \citep{Cho_2003, Skinner_2021_modons, Skinner_2022_cyclogenesis, Kane_2022}, this approach has many advantages. In terms of computational requirements, it scales broadly linearly with the number of regions (here fixed to three), and has no scaling with planetary phases (i.e., time), making it ideal for light-curve modeling. By design, the model already includes the main phase-curve observables (i.e., hotspot offset and hotspot shape). Additionally, since the model is built using the \textsc{TauREx3.1} framework, \textsc{ExPLOR} is automatically compatible with all the \textsc{TauREx3} plugins. \textsc{ExPLOR} can therefore use: GPU accelerated emission and transmission \citep{al-refaie_2021_taurex3.1}, chemical codes \cite[\textsc{ACE, GGchem, FastChem, FRECKLL}: ][]{Agundez_2012, Woitke_2018_GG, Stock_2018, Al-Refaie_2022_FRECKLL}, cloud models \cite[\textsc{YunMa}: ][]{Ma_2023}, stellar activity models \cite[\textsc{ASteRA}: ][]{Thompson_2023}, optimizers \cite[\textsc{MultiNest, UltraNest}: ][]{Feroz_multinest, Buchner_2021}. \textsc{ExPLOR} uses an implementation of the 1.5D phase-curve model to calculate the spectroscopic thermal emission of the planet and its atmosphere along $t$, noted $F_\mathrm{PC}(\lambda, t)$, and the spectroscopic contribution to the planetary radius in transit, noted $\Delta_\mathrm{T}$($\lambda$). Here, $\lambda$ is the wavelength. The radiative transfer calculations from this module do not assume any particular spectral resolution, and are typically ran at the native resolution of the cross-sections, in this work R = 15,000 with {\it ExoMol}\footnote{cross-sections data available at: \url{https://www.exomol.com}} cross-sections \citep{Tennyson_exomol}. \textsc{ExPLOR} also accounts for most of the three-dimensional biases arising from the 3D nature of exoplanet \citep{Feng_2016_inomogeneou, Caldas_2019, Taylor_2020, Changeat_2020_phasecurve1, Zingales_2022, MacDonald_2022, Nixon_2022}.  \\

\underline{\it Modeling transit and eclipse}: To construct the light-curve part of the \textsc{ExPLOR} model (i.e, time/phase conversion, and the transit and eclipse events), we employ the open-source package \textsc{PyLightcurve} \citep{tsiaras_plc}. \textsc{PyLightcurve} allows us to imprint the transit and eclipse events to the planetary emission calculated by the 1.5D \textsc{TauREx} phase-curve model. Combining the time dependent transit light-curve model ($F_\mathrm{T}$) and the normalized eclipse light-curve model ($F_\mathrm{E}'$) from \textsc{PyLightcurve} with the time dependent planetary emission model ($F_\mathrm{PC}$), the final panchromatic light-curve model $F(\lambda, t)$ is given by: 
\begin{equation}
    F(\lambda, t) = F_\mathrm{T}(\Delta_T(\lambda), t) + F_\mathrm{PC}(\lambda, t) \times F_\mathrm{E}'(\lambda, t)
\end{equation}

In this equation, $F_\mathrm{E}'$ is normalized from 0 to 1, as done in \cite{Dang_2018}, since the 1.5D phase-curve model already accounts for the wavelength dependent emission during this event. To compute the transit light-curve drop at each wavelength, we provide the planet-to-star radius ratio (including atmospheric contribution) calculated with the 1.5D phase-curve model (see point above) to \textsc{PyLightcurve}. This means that \textsc{ExPLOR} automatically accounts for the changing apparent size of the planet with wavelengths in both transit and eclipse events \citep{Taylor_2022}. In \textsc{PyLightcurve}, the limb-darkening is modeled by the \cite{Claret_2000} law, for which the coefficients are pre-computed at each wavelength, using for instance an \textsc{ATLAS} grid of stellar models. Note that future instances of \textsc{ExPLOR} could consider retrieval strategies for the limb-darkening coefficients. Here we obtain those coefficients using an automated call to the \textsc{ExoTETHyS}\footnote{\textsc{ExoTHETHyS}: \\ \url{https://github.com/ucl-exoplanets/ExoTETHyS}} package \citep{Morello_2020, Morello_2021_PCBOAT}. \\

The forward model in \textsc{ExPLOR}, when used for a light-curve simulation containing 200 integrations and at R = 15,000 between $\lambda \in [0.3, 50]\, \mu$m, runs in the order of 10 seconds on a single core, meaning that it can be used for sampling with full Bayesian optimizers.

\begin{figure*}
\centering
    \includegraphics[width = 0.9\textwidth]{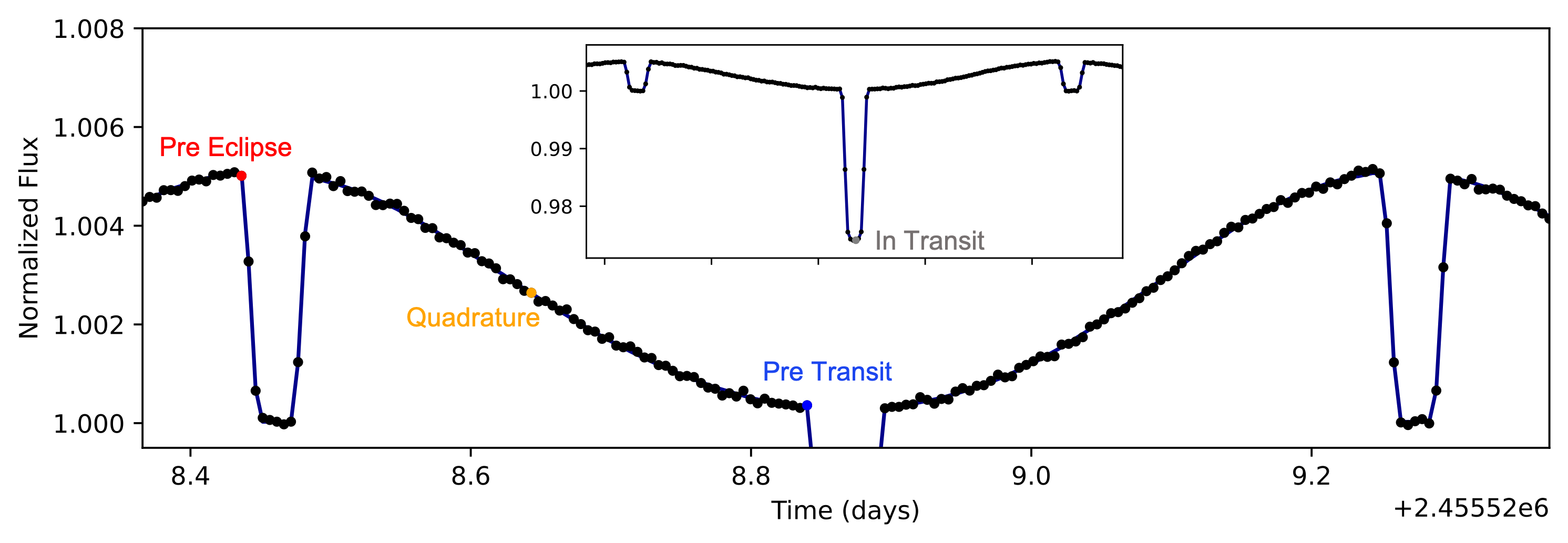}
    \includegraphics[width = 0.9\textwidth]{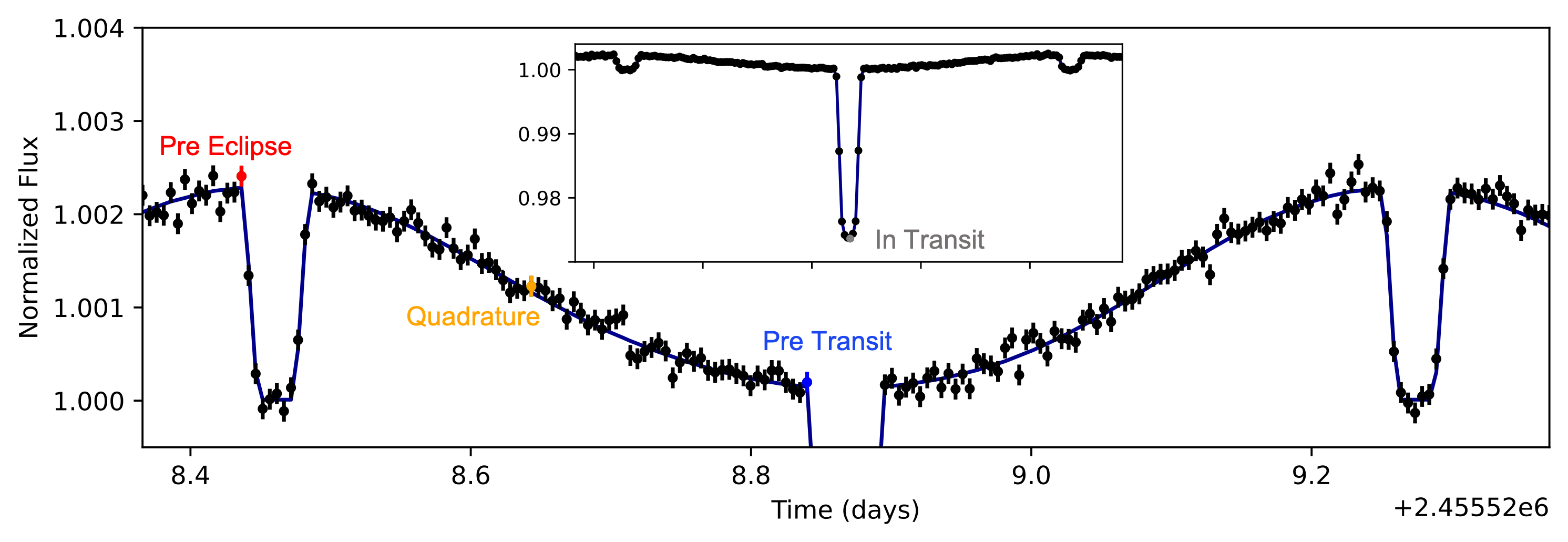}
    \includegraphics[width = 0.49\textwidth]{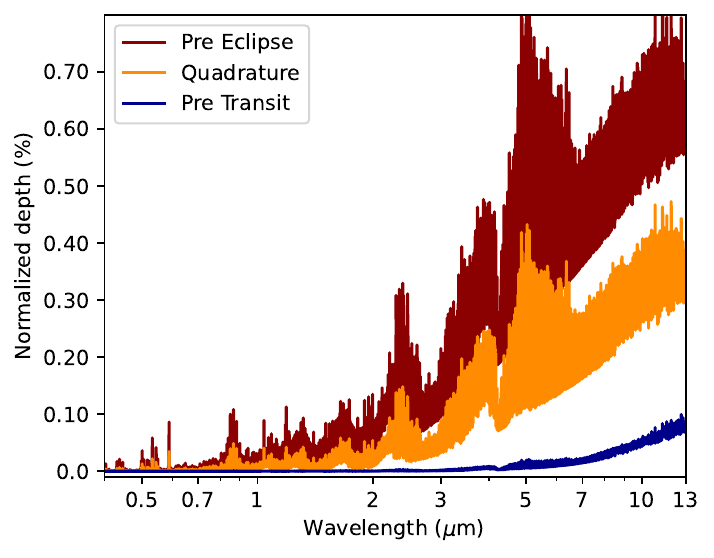}
    \includegraphics[width = 0.49\textwidth]{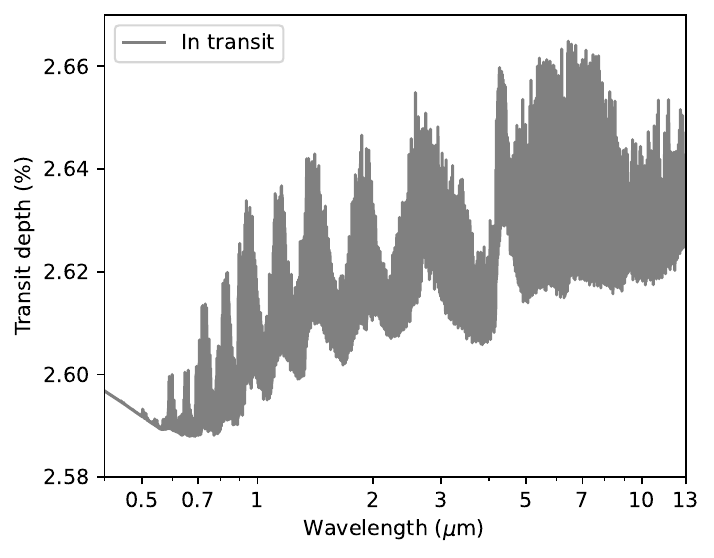}
    \caption{Simulated light-curves for a WASP-43\,b like exoplanet using our panchromatic light-curve model \textsc{ExPLOR}. The top panel shows the white light-curve convolved with the observational noise from JWST-MIRI, and the middle panel shows the same but convolved with the noise profile from ESA-Ariel. The bottom two panels show the underlying simulated spectra (in full resolution) at phases color-marked in the white light-curves (top and middle panels). Note that the top and middle panels are zoomed to illustrate the phase-variations and eclipses, while the insets show the full light-curve with the transit. Each datapoint corresponds to $\Delta t = 7$mn integrations. By definition of the instruments, the JWST-MIRI phase-curve covers $\lambda \in [5, 12]\, \mu$m, while the Ariel phase-curve covers $\lambda \in [0.5, 7.8]\, \mu$m. }
    \label{fig:wasp43_white}
\end{figure*}

\subsection{Applications to a WASP-43\,b like planet}

To validate the \textsc{ExPLOR} panchromatic light-curve strategy, we simulate examples of a WASP-43\,b like exoplanet, as observed by the NASA/ESA/CSA James Webb Space Telescope \cite[JWST:][]{Greene_2016} and the ESA Ariel telescope \citep{Tinetti_2021_redbook}. Our overall methodology is similar to \cite{Changeat_2019_2l}: 1) \textsc{ExPLOR} is used in forward model mode to obtain a high-resolution spectro-temporal map of a phase-curve observation, knowing the true input parameters; 2) the simulated map is convolved to an instrument noise instance for a WASP-43\,b like exoplanet (we consider JWST and Ariel); 3) \textsc{ExPLOR} is used in retrieval mode to recover the free parameters of the model.

\subsubsection{Creation of the simulated observations}

In our example, the planet and star have the same properties as WASP-43\,b \cite[see:][]{hellier_w43}, which is a benchmark target for both observatories \citep{Bean_2018, edwards_ariel, Edwards_2022}. Note, however, that we do not attempt to simulate the real atmosphere of WASP-43\,b. For the forward model, we employ the same setup as in \cite{changeat_2021_phasecurve2}, with the thermal structure of each region being the same as in their Figure 1. More specifically, the thermal structure is parameterized by five temperature -- pressure ($T-p$) nodes for the hotspot and dayside regions, and three $T-p$ nodes for the nightside. The hotspot free parameters (i.e., offset: $\Delta_\mathrm{HS}$, and size: $\alpha_\mathrm{HS}$) are also set to the same values, respectively $\Delta_\mathrm{HS} = -12.3^{\circ}$ and $\alpha_\mathrm{HS} = 40.0^{\circ}$. For the chemistry, we investigate two scenarios showcasing the flexibility of \textsc{ExPLOR}: \\

\underline{\it Scenario 1:} we couple all the regions of the model and assume an hydrogen/helium dominated atmosphere with a solar He/H$_2$ ratio. Trace water and carbon dioxide are then added with respective volume mixing ratios of log(H$_2$O) $= -3$ and log(CO$_2$) $= -4$. \\

\underline{\it Scenario 2:} We simulate the atmosphere at chemical equilibrium using the \textsc{TauREx} plugin: \textsc{ACE}  \citep{Agundez_2012, al-refaie_2021_taurex3.1}. The metallicity (Z) and the C/O ratio in ACE are coupled between the model's regions, but the chemical profiles are not (i.e., each region has a different chemical abundance profile for each molecule). To make this case more realistic, we also include a Grey cloud cover at the nightside using a cloud top pressure of $p = 1000$\,Pa. As fully opaque clouds create the continuum emission, a thermal gradient is required to probe their altitude. To facilitate this, we increase the temperature at the bottom of the atmosphere compared to the previous scenario to $T = 800$\,K. This scenario is a more challenging and realistic case. \\

The radiative transfer calculations include opacities from the molecules H$_2$O \citep{polyansky_h2o}, CO$_2$ \citep{Yurchenko_2020, Chubb_2021_exomol}, CO \citep{li_co_2015}, and CH$_4$ \citep{ExoMol_CH4_new}. Rayleigh scattering \citep{cox_allen_rayleigh} and Collision Induced Absorption by H$_2$-H$_2$ and H$_2$-He pairs \citep{fletcher_h2-h2, abel_h2-h2, abel_h2-he} are also considered. The forward model is ran with 200 phases of equal integration time (here 7 minutes) and at the native resolution of the cross-section (i.e. R = 15,000). The constructed spectro-temporal flux maps cover two eclipses and one transit, including pre and post eclipse baselines. This is the preferred observing phase-curve strategy for most HST and JWST approved programs, and the baseline option for Ariel phase-curves \citep{Tinetti_2021_redbook}.

\subsubsection{JWST and Ariel instrument noise models}

As described previously, the forward model is convolved with an instrument noise model (i.e., of JWST or Ariel) for WASP-43\,b to simulate an observation. For JWST-MIRI, the noise on observation is obtained using the \textsc{ExoCTK Pandexo} tool \citep{Batalha_2017}. For Ariel, the official \textsc{ArielRad} radiometric model \citep{Mugnai_2020} is employed. In the case of JWST, we focus on the Mid Infrared Instrument (MIRI) with the Low Resolution Spectrometer (LRS) as this observation was successfully executed as part of the Early Release Science (ERS) transiting exoplanet program \citep{Bean_2018}. As the telescope noise models were designed with transit and eclipse simulations in mind, the default outputs are the uncertainties on transit and eclipse depths ($\delta T$), assuming half a baseline before and after the event. For independent and normally distributed noise (i.e., ideal instrument behavior), $\delta T$ is related to the noise of individual exposures ($\delta e$) by:
\begin{equation}
    \delta e = \delta T \sqrt{\frac{t_{14}}{2 t_\mathrm{e}}},
\end{equation}
where $t_{14}$ is the transit duration of the exoplanet and $t_\mathrm{e}$ is the integration time of an exposure. This equation is used to obtain the noise on the final spectro-temporal flux maps. Finally, the simulated observations are fed back as input to conduct the panchromatic light-curve retrievals, using \textsc{ExPLOR} in retrieval mode. 

\subsubsection{Retrieval model setup}

\underline{\it Scenario 1: self-retrieval.} We perform a {\it self-retrieval} where the chosen retrieval model has the same assumptions as the forward model. In total, this run has 18 free parameters. We include parameters describing the planet's orbit (mid time: $t_\mathrm{mid}$), the bulk planetary properties (planetary radius: R$_\mathrm{p}$), the atmospheric dynamics (hot-spot offset: $\Delta_\mathrm{HS}$), the atmospheric thermal structure (13 temperature points at fixed pressure labeled with $T$) and atmospheric chemistry. The pressure from the $T-p$ nodes are fixed (for this case to their true value). \cite{changeat_2021_phasecurve2, Rowland_2023} have demonstrated the relevance and computational advantage of this approach. For this scenario, the chemical parameters are the global volume mixing ratios (VMR) of water, log(H$_2$O), and carbon dioxide, log(CO$_2$), fitted from log(VMR) $\in [-12, 0]$. \\

\underline{\it Scenario 2: Uninformed retrieval.} Contrary to Scenario 1, we voluntarily introduce differences between the forward and retrieval models, using an uninformed thermal profiles, to explore a more challenging case. The thermal structure is still parameterized using nodes, but in this case, we arbitrarily select the pressure nodes. We use seven fixed-pressure nodes for the hot-spot and day regions with $p \in \{ 10^6, 10^5, 10^4, 1000, 100, 10, 0.1 \}$\,Pa, and five nodes for the night side with $p \in \{ 10^6, 10^5, 1000, 10, 0.1 \}$\,Pa. Less nodes can be used for the night-side since less information can be extracted from the lower planetary emission. The chemical model is also ACE, with corresponding free parameters being the metallicity (Z) and the C/O ratio, fitted from log(Z) $\in [-2, 2]$ and C/O $\in [0.01, 2]$. In addition we retrieve the cloud top pressure for the night-side region. In total, this is a more extensive retrieval containing 25 free parameters. \\

In both scenarios, we explore the parameter space using non-informative uniform priors and the \textsc{MultiNest} \citep{Feroz_multinest, Buchner_2014} optimizer. We use 250 live points and an evidence tolerance of 0.5.

\section{Results} \label{sec:res}

\begin{figure*}
\centering
    \includegraphics[width = 0.9\textwidth]{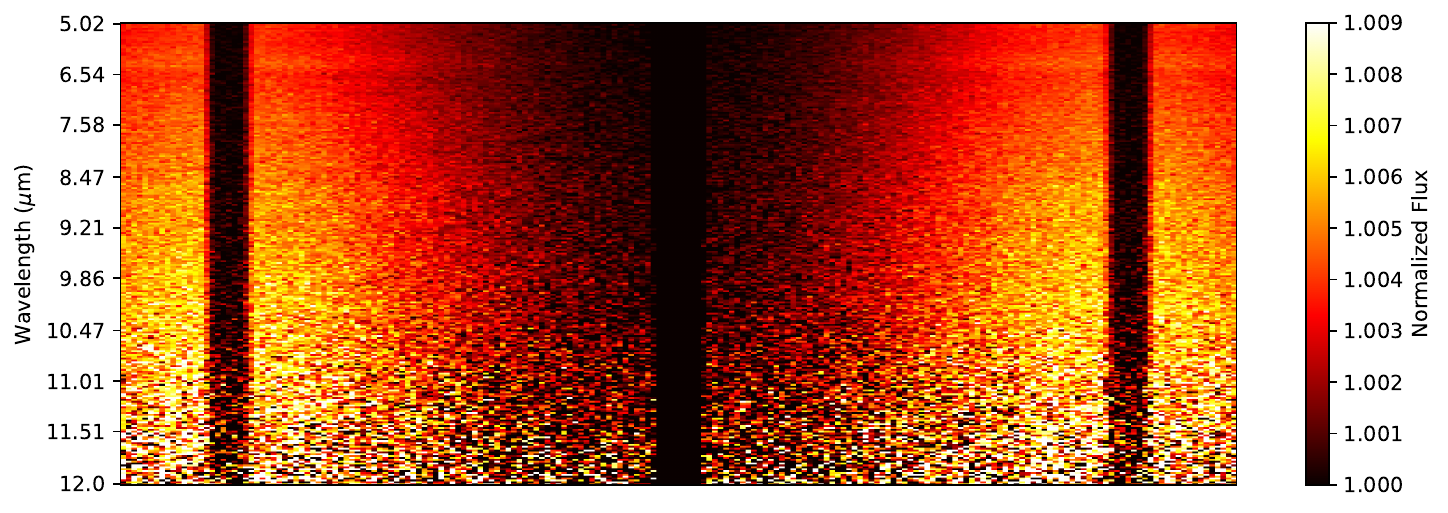}
    \includegraphics[width = 0.9\textwidth]{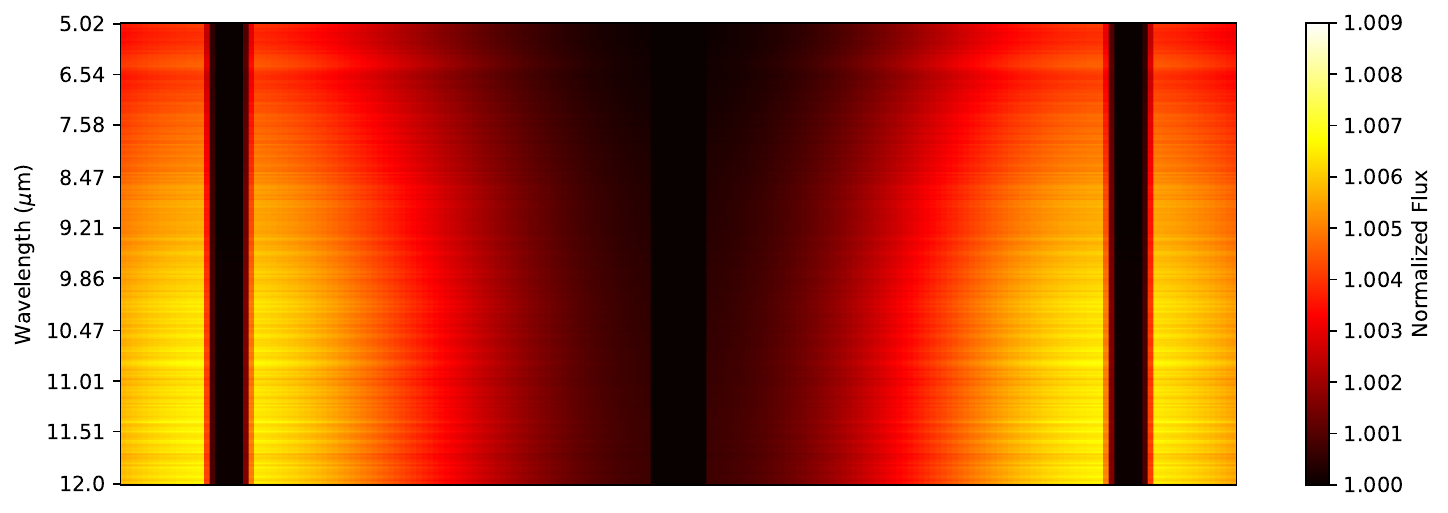}
    \includegraphics[width = 0.9\textwidth]{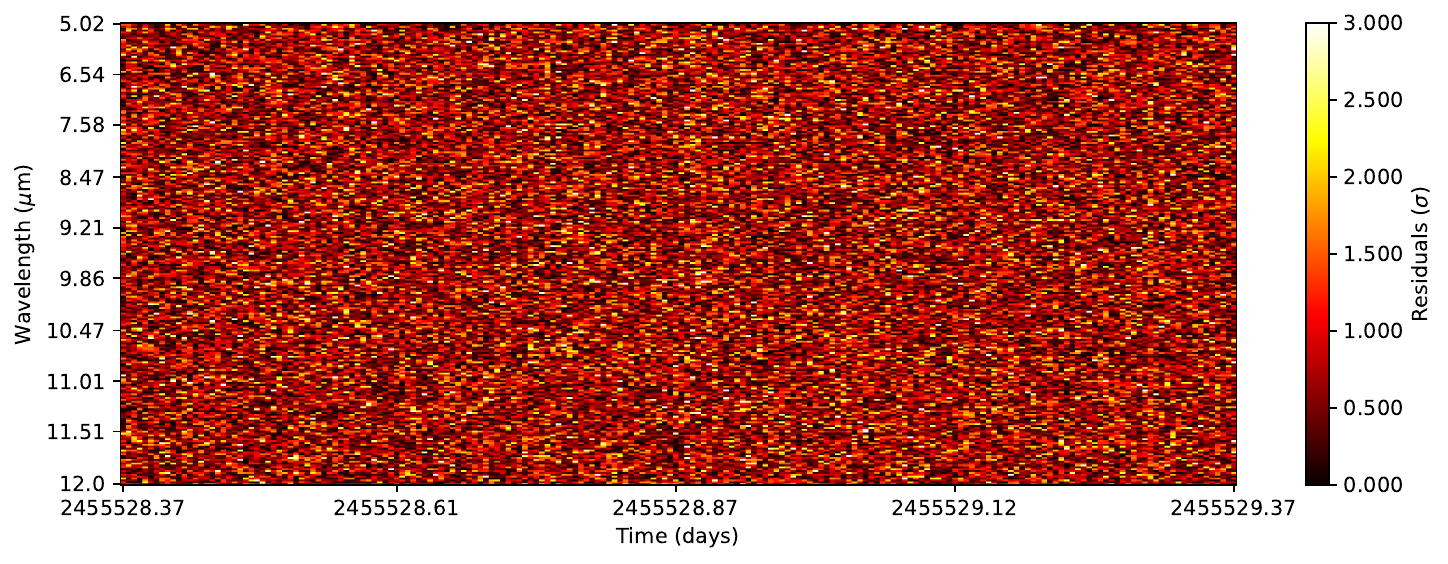}
    \caption{Simulated panchromatic light-curve observation of a WASP-43\,b exoplanet with JWST-MIRI (top) for our Scenario 1. Corresponding best-fit solution from the \textsc{ExPLOR} atmospheric retrieval (middle). Residuals between the simulated observations and the best fit solution and normalized by the observational uncertainties $\sigma$ (bottom). The simulation is created with the same model as shown in Figure \ref{fig:wasp43_white}.}
    \label{fig:wasp43_map_jwst}
\end{figure*}

\begin{figure*}
\centering
    \includegraphics[width = 0.9\textwidth]{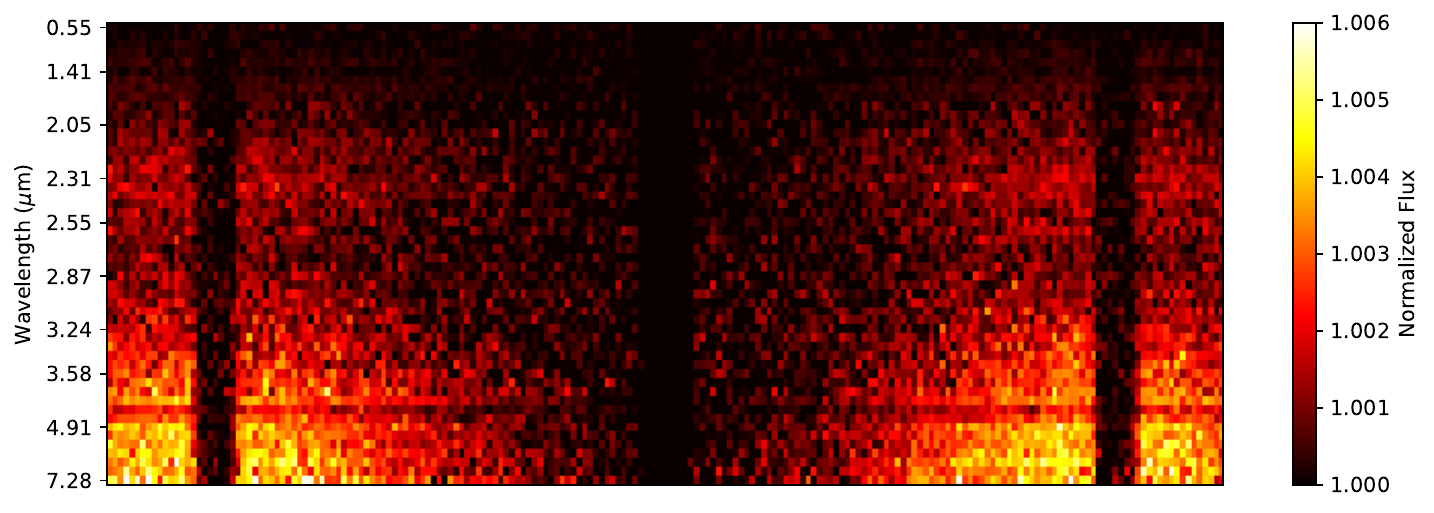}
    \includegraphics[width = 0.9\textwidth]{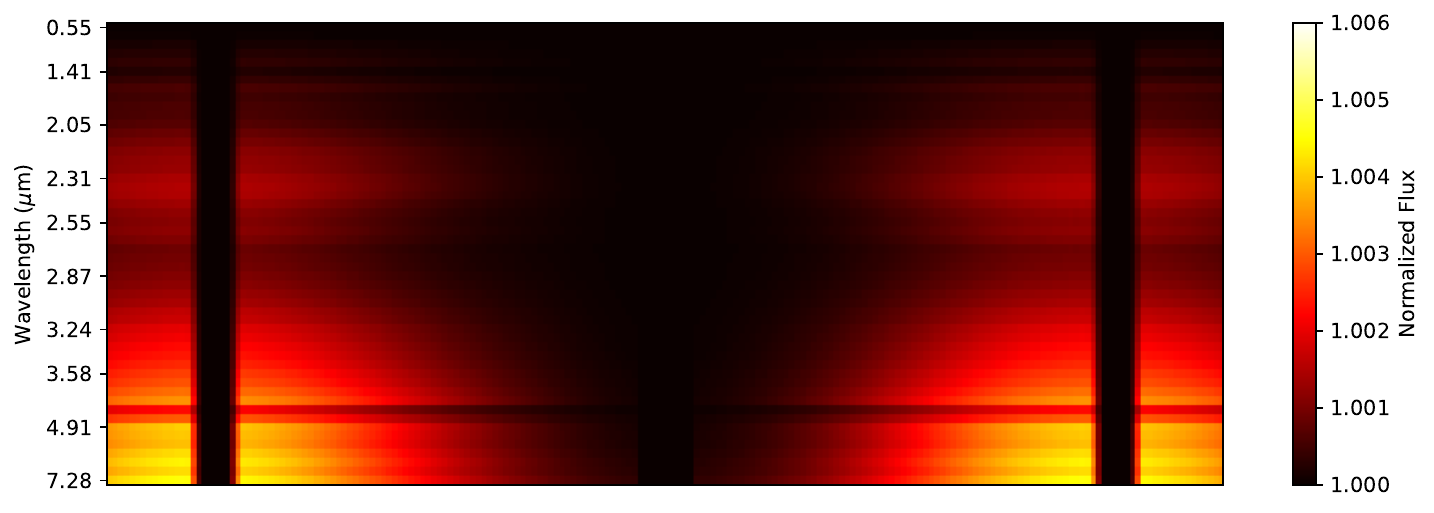}
    \includegraphics[width = 0.9\textwidth]{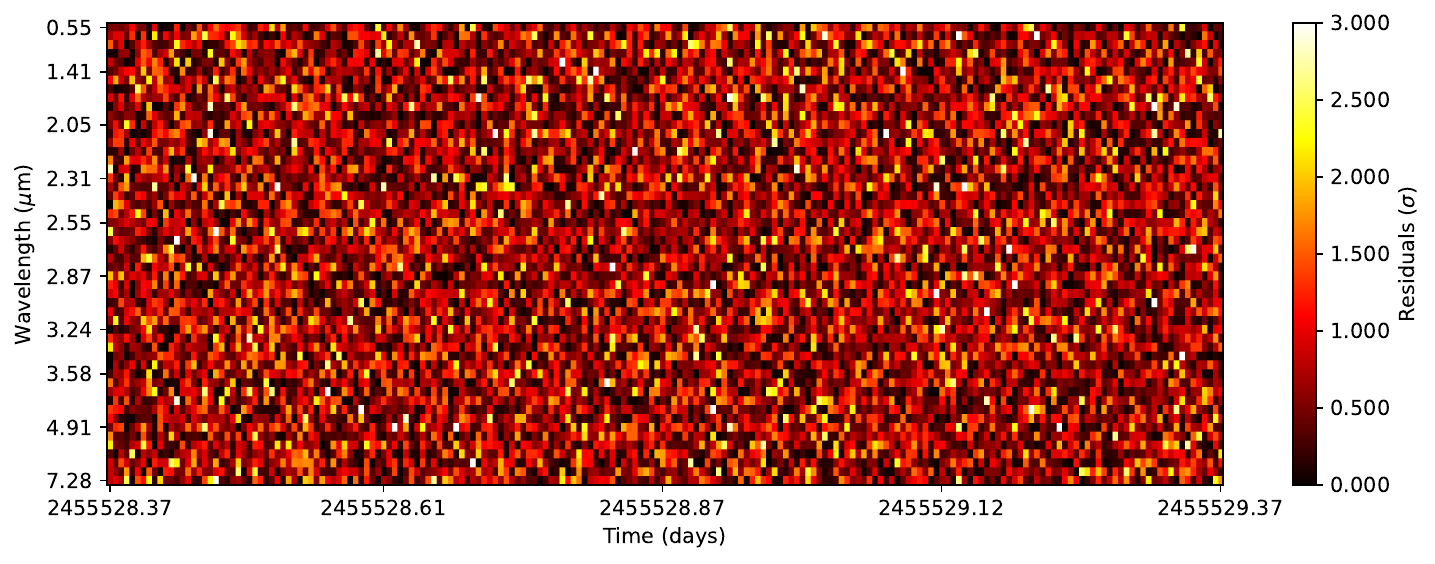}
    \caption{Simulated panchromatic light-curve observation of a WASP-43\,b exoplanet with Ariel (top) for our Scenario 1. Corresponding best-fit solution from the \textsc{ExPLOR} atmospheric retrieval (middle). Residuals between the simulated observations and the best fit solution and normalized by the observational uncertainties $\sigma$ (bottom). The simulation is created with the same model as shown in Figure \ref{fig:wasp43_white}.}
    \label{fig:wasp43_map}
\end{figure*}

With JWST and Ariel mock simulations, we demonstrate the capabilities of \textsc{ExPLOR}. This section focuses on full phase-curve retrievals. More specific examples of eclipse only are discussed in Section \ref{sec:disc}.

\subsection{Scenario 1: Self-retrieval}

For Scenario 1, we show the simulated white light-curves for JWST-MIRI and Ariel and four example spectra at key orbital phases in Figure \ref{fig:wasp43_white}. WASP-43\,b being a benchmark target, those simulations are representative of typical phase-curves for the two observatories. The white light-curves imprints the main dynamical properties of the atmosphere, as shown by the day-night flux contrast and the hotspot offset. In \textsc{ExPLOR}, flux is a function of wavelength and time, so each datapoint on the white light-curve has an underlying spectrum. The spectra (labeled pre-eclipse, quadrature, pre-transit and in-transit) illustrate the spectro-temporal variations of the planet's emission. Entire panchromatic phase-curves are best depicted in {\it spectro-temporal flux maps}, as shown in the top panel of Figure \ref{fig:wasp43_map_jwst} for JWST-MIRI and in the top panel of Figure \ref{fig:wasp43_map} for Ariel. In those maps, horizontal lines of excess/reduced flux correspond to the contribution of radiatively active sources. For instance, H$_2$O induces oscillating flux dimming around $\lambda \in \{ 2, 3, 7 \}\, \mu$m via its absorbing properties, while CO$_2$ is responsible for the much narrower absorption band at $\lambda = 4.5\, \mu$m. Such maps are the input of \textsc{ExPLOR} retrievals. In our tests, the simulated observations are not binned in wavelength and are passed to \textsc{ExPLOR} at the native instrument resolution. While some of the JWST-MIRI detector systematics appear to be mitigated by spectral binning \citep{Bouwman_2023}, noise properties are generally best handled at native resolution due to correlated noise \cite{Espinoza_2023, Holmberg_2023}. In practice, the simulations have 200 time steps with the JWST-MIRI observation having 247 wavelength channels\footnote{wavelengths $\lambda > 12\, \mu$m were cut due to their much higher observational noise} and the Ariel observation having 52 wavelength channels.

Performing the self-retrievals as described in Section \ref{sec:meth}, we obtain the best-fit spectro-temporal flux maps in the middle panels of Figure \ref{fig:wasp43_map_jwst} (JWST-MIRI) and Figure \ref{fig:wasp43_map} (Ariel). The residuals (bottom panels) are consistent with white noise, as expected in a self-retrieval exercise. We show the extracted $T-p$ structures for the three simulated regions (hotspot, dayside and nightside) in Figure \ref{fig:wasp43_tp}, which are consistent with the input true profiles. The 1.5D model in \textsc{ExPLOR}, combined with the highly informative panchromatic light-curve approach, allows us to infer detailed thermal structures in those WASP-43\,b simulations. The chemistry is also accurately constrained as shown by the full posterior distributions in Figure \ref{fig:wasp43_corner_JWST}. Note that such high level of information extraction is made possible by bypassing the construction of reduced spectra, which is required for conventional atmospheric retrievals (see later discussion). From the posterior distributions, we also demonstrate that orbital, bulk and atmospheric (dynamical, thermal, and chemical) information can be efficiently extracted from the panchromatic light-curves, thus avoiding the parameter conditioning paradigm emphasized in \cite{Yip_2020_LC}. In this WASP-43\,b simulation, the observations do not probe well the fully opaque deep layers of the atmospheres, leading to larger uncertainties on the values of $T_\mathrm{surf}$, representing the temperatures at the surface ($p = 10$\,bar), as expected. Overall, this example demonstrate the relevance of panchromatic light-curve retrievals.

\subsection{Scenario 2: Uninformed retrieval}

A similar exercise is reproduced considering a more complex cloudy atmosphere at chemical equilibrium. For simplicity, only the JWST-MIRI case is explored but similar results should be achievable with Ariel. As explained, the retrieval model is slightly different to the forward model (i.e, this is not a self-retrieval), making this case more realistic.

As with Scenario 1, we demonstrate that the retrieval recovers a relevant interpretation of the simulated atmosphere. The spectro-temporal flux maps for this scenario are shown in Figure \ref{fig:wasp43_map_ggchem} and the corresponding retrieved thermal structure is shown in Figure \ref{fig:wasp43_tp} (right panel). Due to the added complexity of the model (i.e., higher dimensionality), this case has a much higher computing requirement than the self-retrieval described above. This is also because the plugin \textsc{ACE} \citep{Agundez_2dchemical_HD209_HD189} is not GPU-enabled, thus reducing the performance gain from the GPU-enabled phase-curve emission model. The recovered posterior distributions for this scenario are shown in Figure \ref{fig:wasp43_corner_JWST_ggchem}, showing the correlation between the 25 free parameters of our problem. Importantly, note that there is no true value for the $T-p$ nodes of this retrieval since the node pressures are voluntarily made different between forward and retrieval models. Reasonable reference values are instead provided by interpolating the temperature from the forward model at the relevant pressures. Importantly, we find that the introduction of slight modeling differences between the forward and retrieval would not affect the interpretation of this atmosphere, showing robustness and stability. 

As opposed to more standard atmospheric retrievals on reduced spectra, fitting directly the spectral light-curves marginalizes over the full set of relevant parameters, allowing us to ensure a consistent propagation of the covariance between each parameter. By interpreting the ``un-processed'' data directly, our panchromatic light-curve retrieval also allows us to maximize the information content and get more precise estimates of the planetary atmosphere. This is shown by the particularly tight constraints that can be obtained on the free parameters by using this method.

\section{Discussion} \label{sec:disc}

\subsection{Why should we care about time in spectral retrievals?}

\begin{figure}
\centering
    \includegraphics[width = 0.48\textwidth]{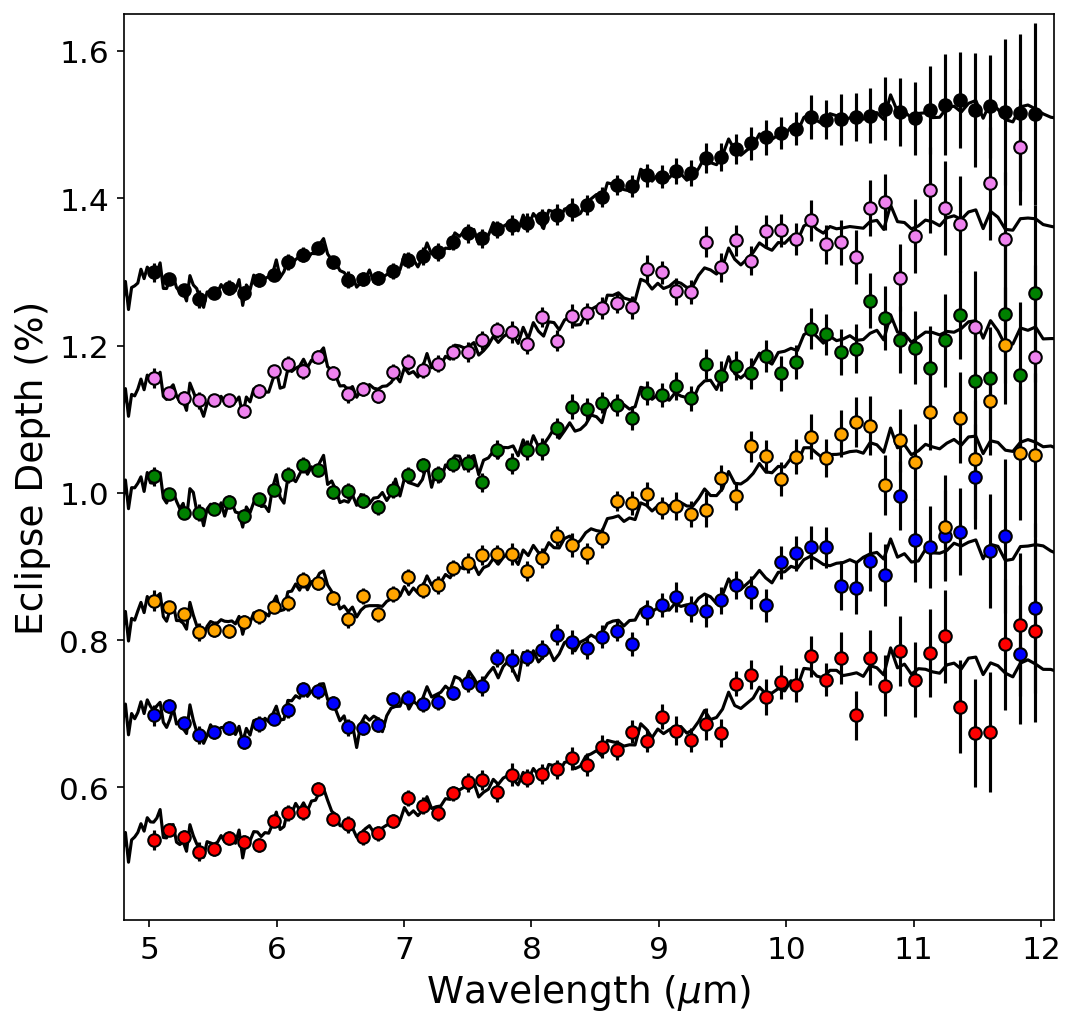}
    \caption{Observed spectra obtained when the \textsc{ExPLOR} simulations are reduced using \textsc{PoP} (datapoints), and best-fit spectra (solid lines) from corresponding 1D \textsc{TauREx3} retrievals. The reduction was performed on simulated eclipses of WASP-43\,b that are similar to the Scenario 1 (the raw data shown in Figure \ref{fig:test_pure}). The black case corresponds to an un-scattered spectro-temporal flux map, while the colored cases are five different instances of scattered spectro-temporal flux maps. Note that the spectra are offset for better clarity. Overall the 1D \textsc{TauREx3} retrievals are able to explain the data.}
    \label{fig:wasp43_spectra_fit}
\end{figure}

\begin{figure*}
\centering
    \includegraphics[width = 0.36\textwidth]{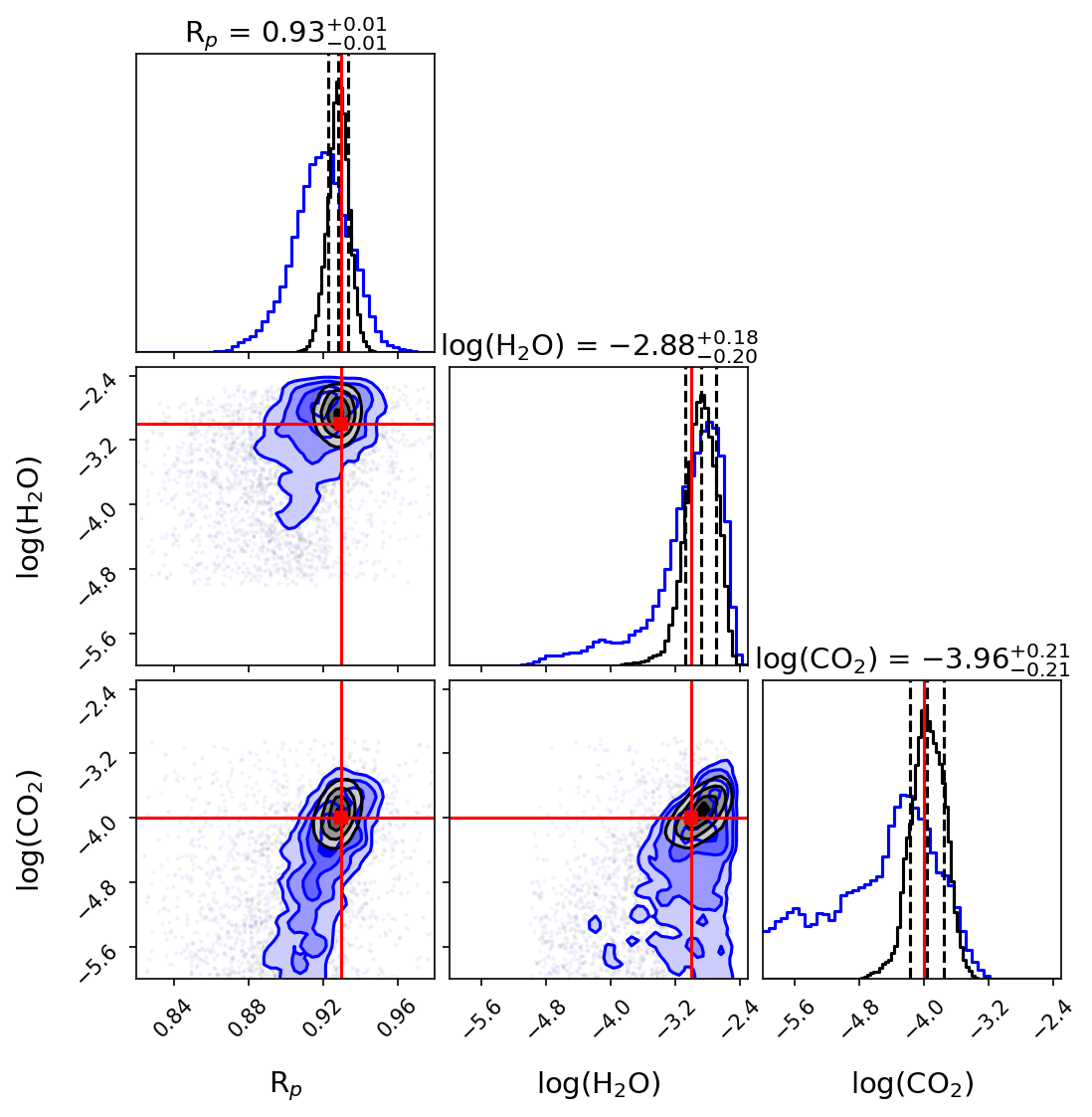}
    \includegraphics[width = 0.36\textwidth]{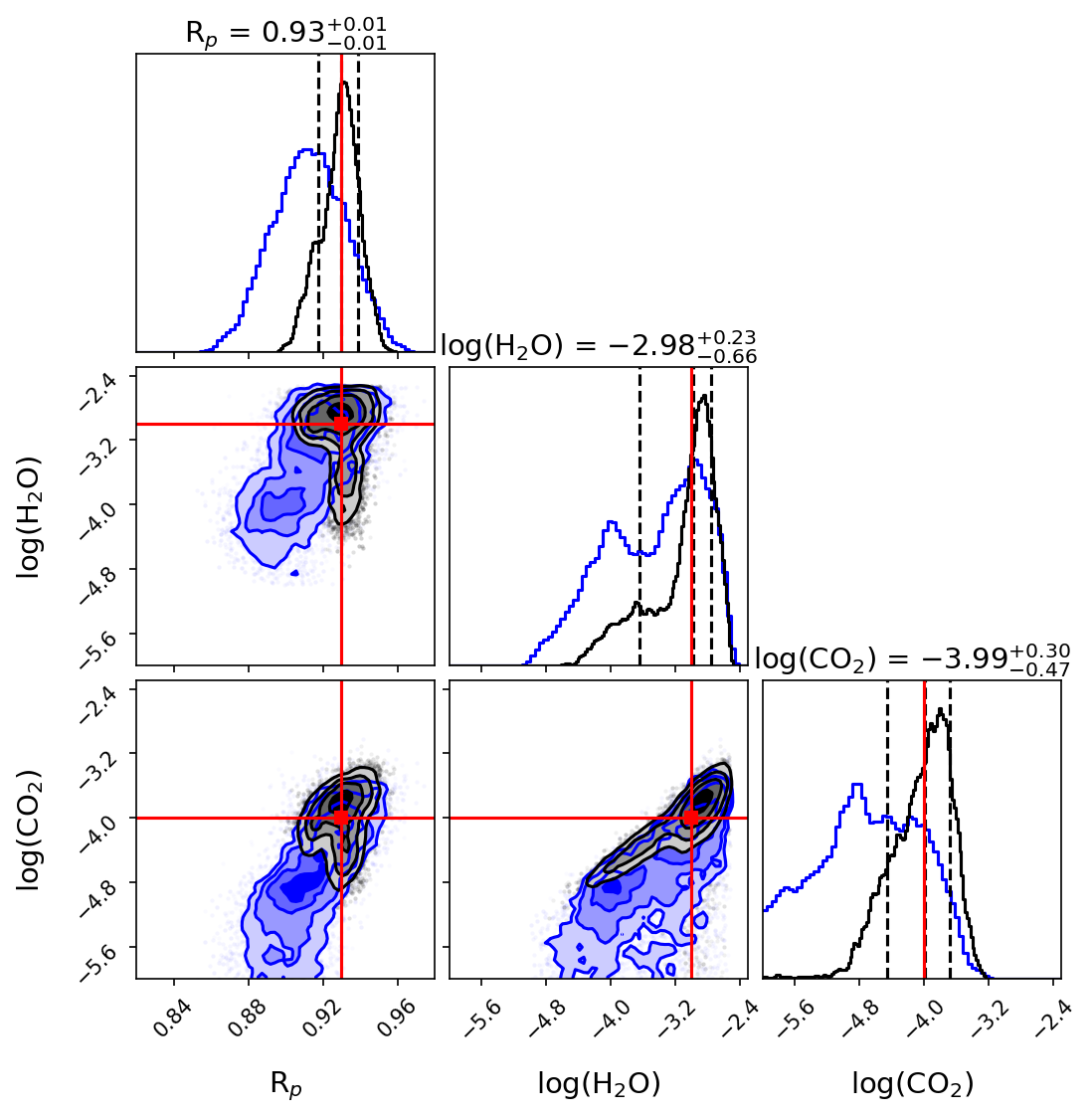}
    \includegraphics[width = 0.26\textwidth]{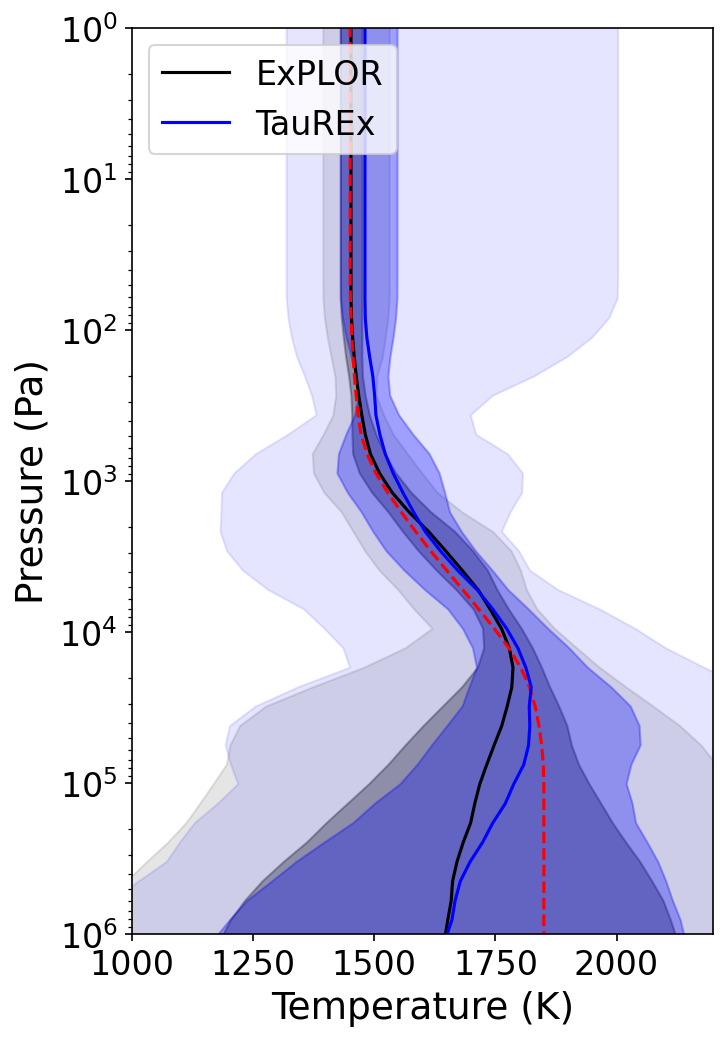}
    \caption{Corner plots (left: un-scattered input; middle: combination of five independently scattered inputs) from the \textsc{ExPLOR} (black) and the \textsc{TauREx} (blue) retrievals obtained for the 1D (i.e, spatially homogeneous atmosphere) JWST eclipse examples, and retrieved $T-p$ structure (right) for the un-scattered case. In the scattered case the final posteriors are obtained by adding the re-sampled weighted traces from the five different scattered instances. In the $T-p$ plot, the solid lines and shaded regions are median, 1$\sigma$ and 3$\sigma$ regions, while the dashed lines are the true values from the forward model. Widening of the posteriors and the retrieved $T-p$ profiles for the \textsc{TauREx} retrievals, even when the maps are un-scattered, shows the effect of information dilution arising from the reduction process. In the scattered case, the \textsc{TauREx} solutions are less stable to the noise properties, most likely due to incorrect propagation of non-gaussian components arising during the reduction process. }
    \label{fig:wasp43_spectra_corners}
\end{figure*}

The atmosphere of transiting exoplanets is primarily characterized using spectroscopic time-series obtained around transit and eclipse events. Those events are important because they isolate the planetary signal from the stellar signal. Time is a key element in those observations, but it is rarely employed at the retrieval stage. As previously said, the time-series are usually first cleaned from the instrument systematics and compressed into a planetary spectrum, using data reduction pipelines. For instance the open source codes \textsc{Iraclis} \citep{tsiaras_hd209}, \textsc{CASCADe} \citep{Lahuis_2020}, or \textsc{Eureka!} \citep{Bell_2022_eureka} transform raw HST/JWST images into a single processed spectrum by fitting a simplified light-curve model (usually convolved with some instrument systematics model). The processed spectrum, however, remains a form of summary statistics or real-valued statistics (i.e., mean, median and variance). In other words, the raw observation is compressed to a series of Gaussian distribution with a vector of means and standard deviations (i.e., a multivariate Gaussian distribution with diagonal elements only). Here, we discuss how the astrophysical signal (i.e., the exo-atmospheric information) is propagated during those reduction steps. We illustrate this using a controlled scenario restricted to the eclipse of a WASP-43\,b like planet. We briefly describe our approach here, but more details are available in Appendix B. Ultimately, our goal is to compare retrievals performed at the light-curve level (i.e., referred as \textsc{ExPLOR} retrievals) against retrievals performed at the spectrum level (referred as \textsc{TauREx} retrievals). Following the standard practice of the literature, we obtain an eclipse spectrum from the \textsc{ExPLOR} spectro-temporal flux map by fitting a simplified light-curve eclipse model with \textsc{PoP} \citep{Changeat_2023}. The resulting spectrum is then interpreted using the standard 1D retrievals of \textsc{TauREx3} and compared against the \textsc{ExPLOR} retrieval. Our analysis is done for an un-scattered spectro-temporal flux map, and reproduced for five independently scattered instances (i.e., the spectro-temporal flux map is normally scattered according to the JWST-MIRI instrument noise). To ensure fair and realistic comparison, we produce the spectro-temporal map of a 1D atmosphere (i.e, spatial variation are not included) using a our Scenario 1 slighly modified. We set the $T-p$ profile to the hotspot values everywhere in the planet. A more subtle case with spatial variations is presented in Appendix B, showing similar conclusions. For our 1D case, the input and \textsc{ExPLOR} best-fit maps are shown in Figure \ref{fig:test_pure} while the reduced spectra and best-fit \textsc{TauREx3} retrievals are shown in Figure \ref{fig:wasp43_spectra_fit}). The comparison demonstrates the relevance of the literature approach (i.e, conventional spectral retrievals can interpret the input data), but also illustrates the compression of information from the 2D spectro-temporal maps to the 1D spectra. The posterior distributions in Figure \ref{fig:wasp43_spectra_corners} further reinforce this point, quantifying the imperfect propagation of the astrophysical signal and the noise properties as shown by the much broader distributions. CO$_2$ for instance, which is consistently (i.e., in all the tested scattered instances) retrieved with \textsc{ExPLOR} is often missed by the \textsc{TauREx} retrievals.

From a theoretical point of view, working on summary statistics has major drawbacks, which are summarized here: \\

$\bullet$ {\it Information dilution:} As the extracted spectrum is constructed from summary statistics, it does not encode the full information content of the observations. This is shown explicitly by the posterior distributions (Figure \ref{fig:wasp43_spectra_corners}), which have larger uncertainties and lower level of detection significance (see CO$_2$ posteriors) when performing spectral retrievals (\textsc{TauREx} retrievals). An obvious example of such information dilution is also the time-dependent information available during eclipse, which allows us to perform eclipse mapping (see additional case in Figure \ref{fig:test_pure_temperature}). \\

$\bullet$ {\it Noise propagation:} The underlying assumption used to construct the observed spectra is to assume Gaussian error propagation. Observational noise and light-curve models have some non-Gaussian components that cannot be well handled by summary statistics \citep{Yip_2020_LC}. For instance, when reduction pipelines retrieve the planetary radius from each light-curve independently, the posterior distribution is not always Gaussian, but often spectral retrievals only handle mean and variance for each wavelength. Additionally, when performing data reduction, each light-curve is considered as an independent measurement meaning that their covariance is assumed to be zero. This is not expected to be true, especially when instrument systematics are present. Noise propagation along the reduction chain is imperfect in our example of scattered data, as shown by Figure \ref{fig:wasp43_spectra_corners}.  \\

$\bullet$ {\it Parameter conditioning:} Since the standard retrieval techniques do not marginalize over all the parameters of the problem (i.e. they do not explicitly include instrument systematics and orbital configuration), the extracted parameters could be biased or their uncertainties under-estimated. This is because the solution is conditioned over a particular instance (rather than the full probability distribution) as the fitting of the instrument and the orbital dynamics is de-coupled at the reduction stage. We do not explore this point more, but the panchromatic light-curve approach should naturally avoid parameter conditioning by exploring the relevant parameter space.\\

In addition to solving those issues, the introduction of time to retrievals offers several advantages: \\ 


$\bullet$ {\it It is more flexible:} This method allows to naturally model time-varying phenomena. By resolving the ingress and egress in transit and eclipse, this allows eclipse mapping and could also be used to constrain patchy clouds (i.e, when the east and west terminator limbs are different). Inspection of residuals could help uncover and characterize stellar activity, spot crossing events, chemical inhomogeneities (i.e, disequilibrium chemistry). \\

$\bullet$ {\it It is a more natural approach to model complex observations:} It can handle complex situations with minimum modeling efforts. For instance, modeling and retrieval of transits, eclipses and phase-curves involving multiple planets of the same system, or from observations of the same planet using different techniques, is trivial. 

\subsection{Note on computational requirements}

Recently, the analysis of revolutionizing JWST data has challenged the scalability and efficiency of atmospheric retrieval methods, which by including more accurate and complex processes require significantly more computational resources. An important barrier encountered by modern retrievals of the JWST era -- on top of the increase in the forward model compute time -- regards the wider prior space. When explored using mainstream sampling-based techniques, the prior space can easily span 30+ free parameters, reaching the {\it curse of dimensionality} \citep{Buchner_2021}. This is particularly the case for {\it more-than-1D} retrievals (i.e., like \textsc{ExPLOR}) requiring additional parameterizations for the spatial and temporal dimensions. Additional limitations can arise when moving away from {\it clean} simulated datasets. Real observations display strong spectro-temporal instrument systematics \cite[see e.g.,][]{Bell_2023, Kempton2023} or require improved stellar properties \cite[i.e, accurate limb-darkening properties][]{Rustamkulov_2023} that need to be modeled simultaneously, potentially adding several degrees of freedom to the problem.
On one side, potential avenues to overcome those issues include the development of more physically motivated retrievals (i.e, with reduced degrees of freedom) based on our continuously evolving understanding of exo-atmospheres and instruments. On the other side, novel inference techniques, either based on more scalable sampling-based methods -- such as dynamic nested sampling \citep{Higson_2019} or Phantom-powered nested sampling \citep{Albert_2023} -- or alternative techniques -- such as Variational Inference \citep{Yip_2022}, Simulation Based Inference \citep{Vasist2023, Gebhard} or ML-accelerated surrogate modelling \citep{Himes_2022} -- needs to be developed.

For reference, our \textsc{ExPLOR} retrieval of the un-scattered eclipse case of Figure \ref{fig:wasp43_spectra_corners} utilizes 230.4 CPUh (about 111\,mn, for 61,000 evaluations) using two AMD EPYC 7543 nodes (total of 128 cores) when a \textsc{TauREx} retrieval of the derived spectrum uses 9.2 CPUh (about 5\,mn for 17,000 evaluations) on the same machines. While direct comparison is difficult since the \textsc{ExPLOR} run includes an additional free parameter for the mid-transit time and models the planet in 1.5D, this difference clearly demonstrates the extent of modern computational challenges for retrievals.

\section{Conclusion} \label{sec:conc}

We introduce an innovative retrieval method designed to infer atmospheric properties from panchromatic light-curves directly. This method, embedded in the novel \textsc{ExPLOR} model, operates closer to the data as compared to conventional retrieval strategies acting on spectra, providing several advantages. In particular, the noise properties of the observations can be more directly integrated into the retrieval process, avoiding issues such as information dilution, loss of noise characteristics and parameter conditioning. By treating time in the atmospheric model, this generalized approach also offers a more flexible and natural strategy for modeling observations of transiting exoplanets. In particular, \textsc{ExPLOR} can model transit and eclipse situations, but also phase-curves, eclipse mapping and other more complex situations. In this study, we conduct mock simulations to validate our model implementation and showcase its capabilities. Overall, this work lays the groundwork for the development of approaches that are more intimately connected to observational data, which will ultimately lead to a more efficient use of telescope data. In future investigations, we will more directly compare the performances of panchromatic light-curve retrievals with the standard approach and apply our model to real observations obtained by next-generation space-based facilities.

\section*{Data and materials availability}

The data and software that support the findings of this study can be made available upon request from the authors.

\section*{Acknowledgements}

The authors wish to thank Prof. Tinetti and Prof. Waldmann for providing their expert opinion and useful comments on the results of this project. We also thank the anonymous reviewer for their insightful comments and recommendations.

QC is recipient of the 2022 European Space Agency Research Fellowship. This research was supported by the European Research Council, Horizon 2020 (758892 ExoAI) and the STFC/UKSA Grant ST/W00254X/1. Part of this work was also supported by NAOJ Research Coordination Committee, NINS, NAOJ-RCC-2301-0401. YI was also supported by JSPS KAKENHI grant No. 22K14090.

We acknowledge the availability and support from the High-Performance Computing platforms (HPC) from the Simons Foundation (Flatiron), DIRAC, and OzSTAR, which provided the computing resources necessary to perform this work. This work utilised the Cambridge Service for Data-Driven Discovery (CSD3), part of which is operated by the University of Cambridge Research Computing on behalf of the STFC DiRAC HPC Facility (www.dirac.ac.uk). The DiRAC component of CSD3 was funded by BEIS capital funding via STFC capital grants ST/P002307/1 and ST/R002452/1 and STFC operations grant ST/R00689X/1. DiRAC is part of the National e-Infrastructure. This work utilised the OzSTAR national facility at Swinburne University of Technology. The OzSTAR program receives funding in part from the Astronomy National Collaborative Research Infrastructure Strategy (NCRIS) allocation provided by the Australian Government. 



\bibliographystyle{aasjournal}
\bibliography{main}



\renewcommand\thesection{\Alph{section}}
\renewcommand\thesubsection{\thesection.\arabic{subsection}}
\value{section} = 0

\section{Appendix A: Complementary Figures}

\setcounter{figure}{0}
\renewcommand{\thefigure}{A\arabic{figure}}
\renewcommand{\theHfigure}{A\arabic{figure}}

\begin{figure}[H]
\centering
    \includegraphics[width = 0.27\textwidth]{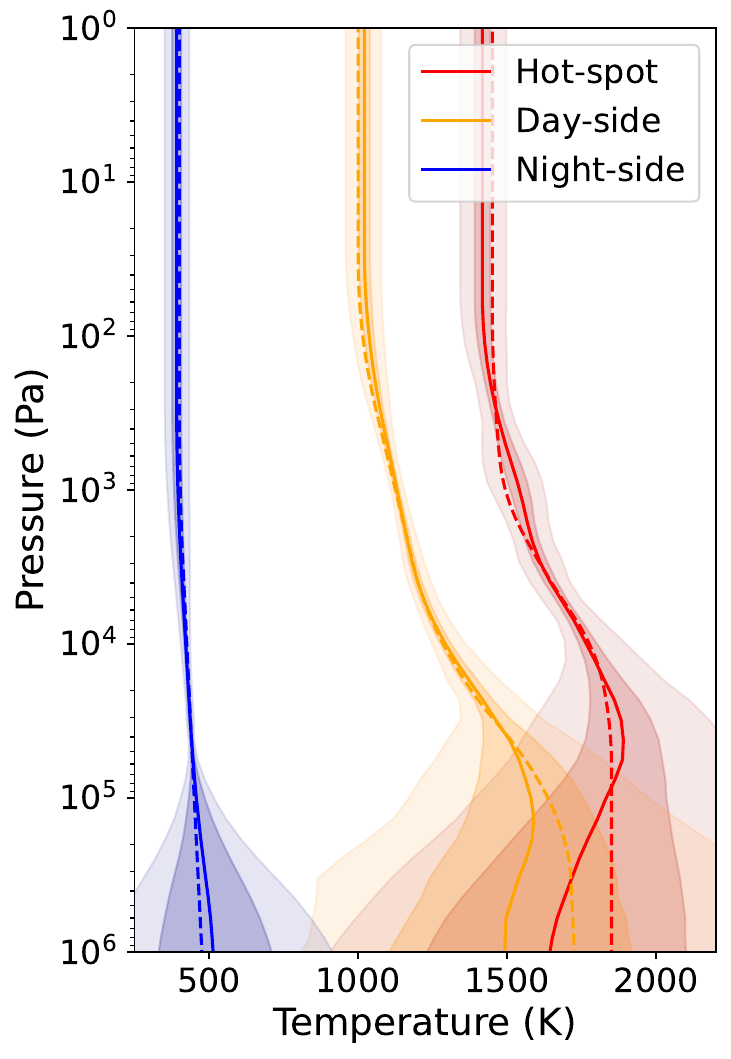}
    \includegraphics[width = 0.27\textwidth]{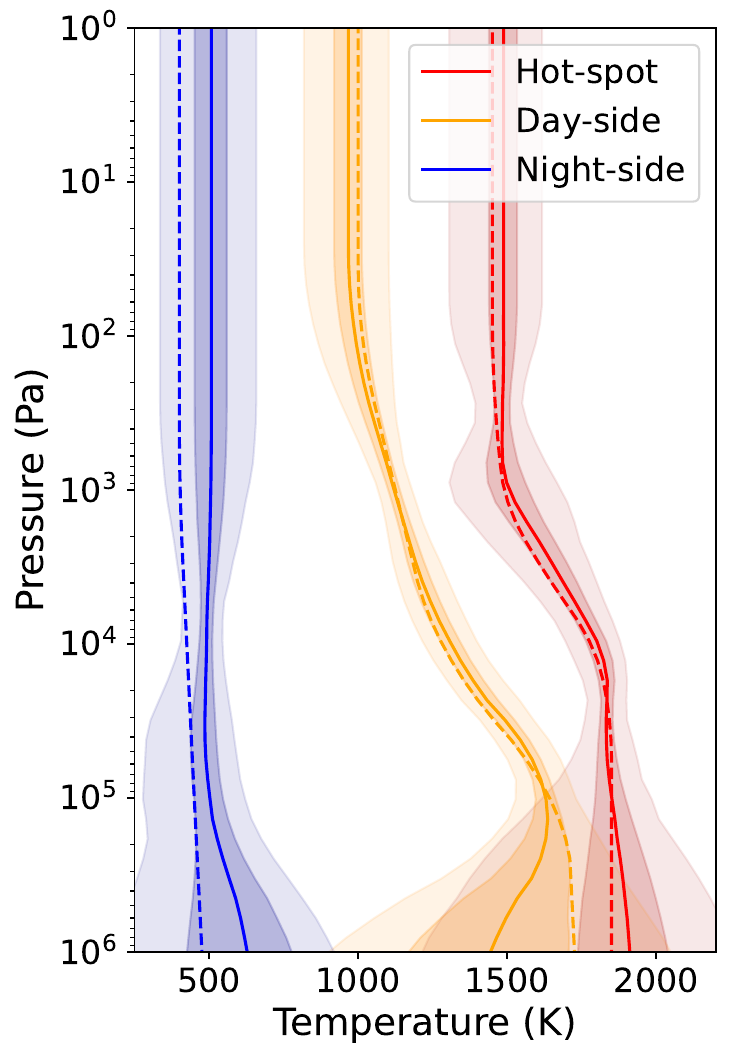}
    \includegraphics[width = 0.27\textwidth]{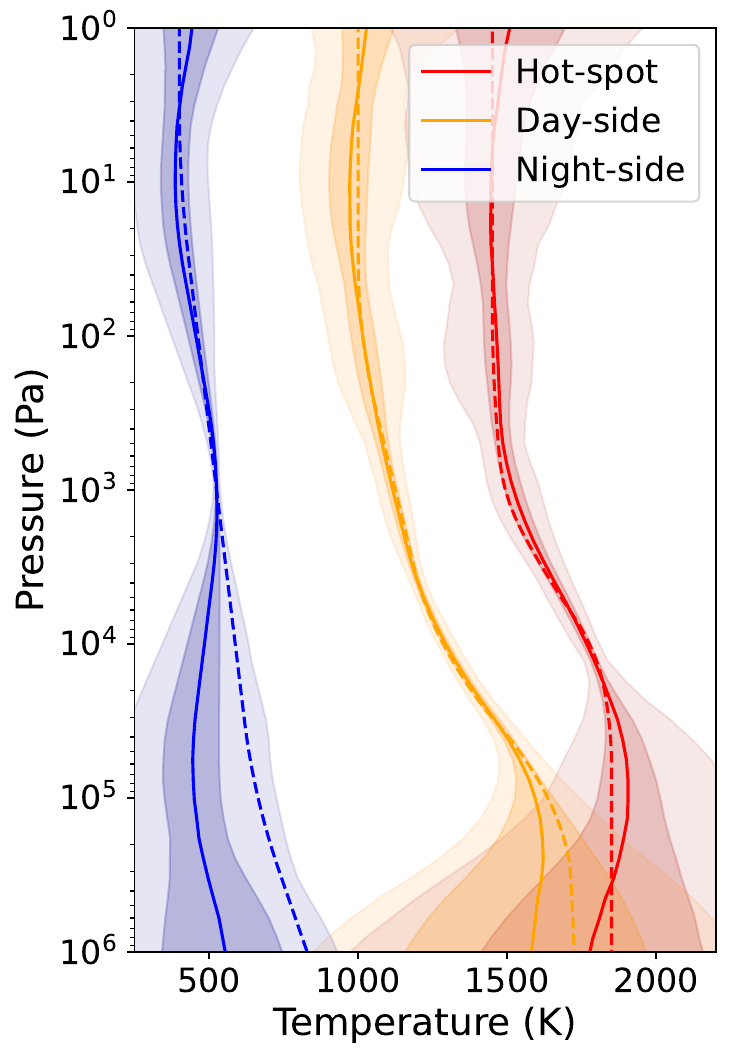}
    \caption{Retrieved temperature structure for the WASP-43\,b panchromatic phase-curve simulations. Left: JWST-MIRI simulation for Scenario 1. Middle: Ariel simulation for Scenario 1. Right: JWST-MIRI simulation for Scenario 2. The solid lines and shaded regions are median, 1$\sigma$ and 3$\sigma$ regions, while the dashed lines are the true values from the forward model. In all three simulations, the panchromatic light-curve retrievals manage to infer the thermal structure of the planet for all three regions.}
    \label{fig:wasp43_tp}
\end{figure}

\begin{figure}[H]
\centering
    \includegraphics[width = 0.99\textwidth]{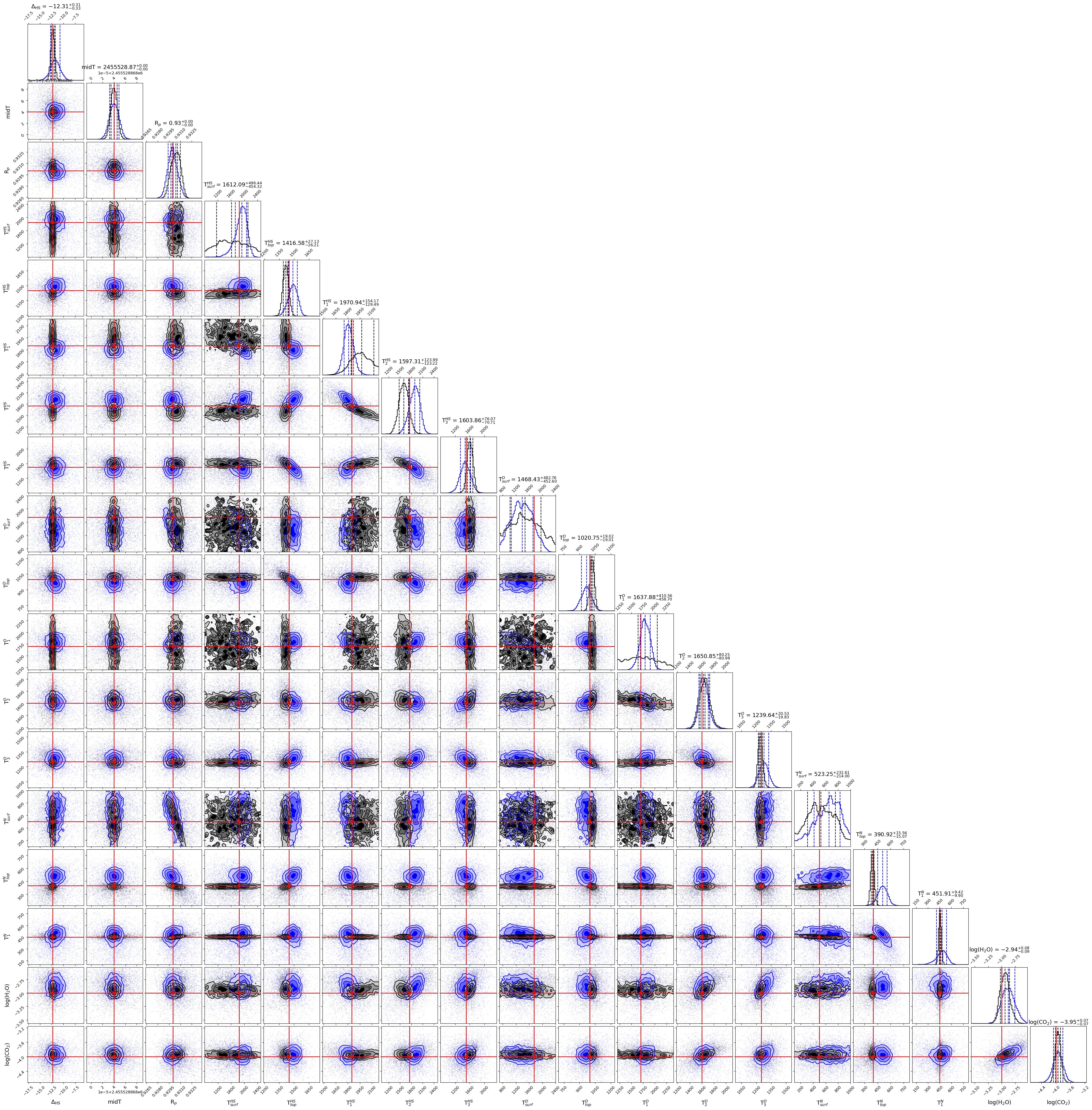}
    \caption{Full corner plot of the \textsc{ExPLOR} panchromatic light-curve retrieval on the WASP-43\,b like case for Scenario 1. This shows simulations for JWST-MIRI (black) and Ariel (blue) telescopes. The true value of this simulation is indicated by the red crosses.}
    \label{fig:wasp43_corner_JWST}
\end{figure}

\begin{figure*}
\centering
    \includegraphics[width = 0.9\textwidth]{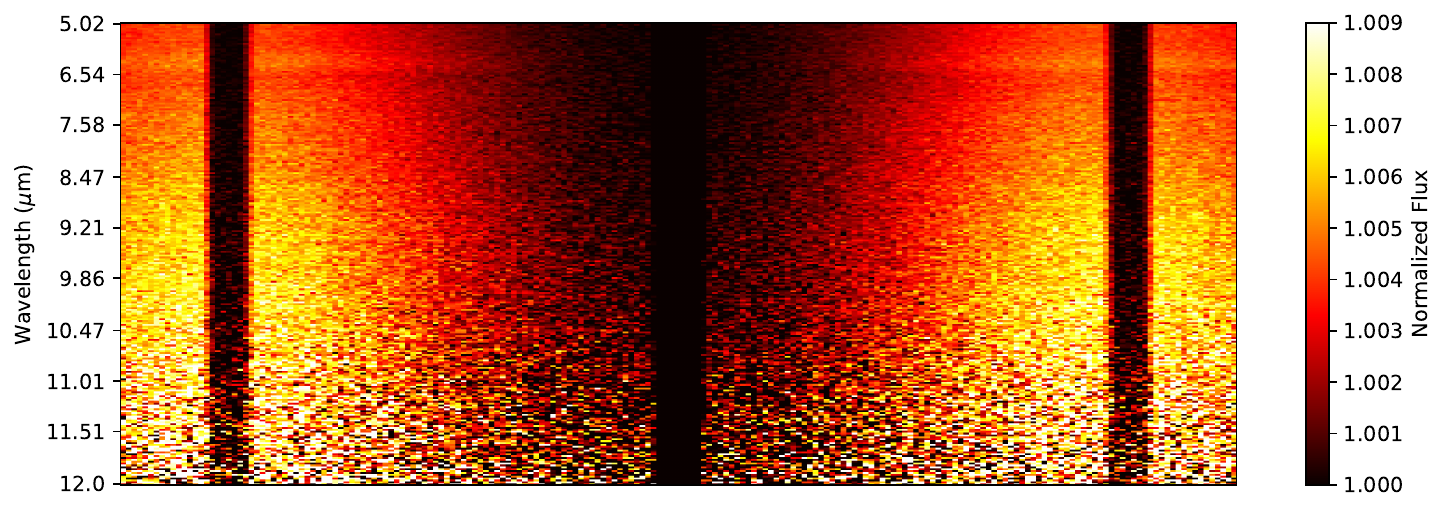}
    \includegraphics[width = 0.9\textwidth]{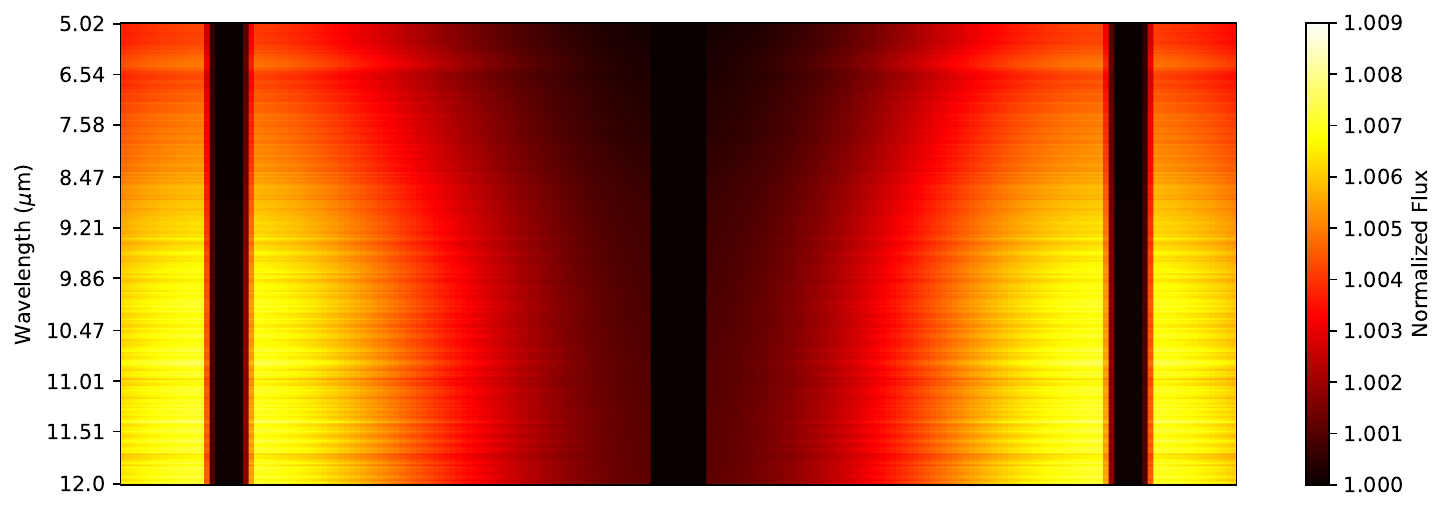}
    \includegraphics[width = 0.9\textwidth]{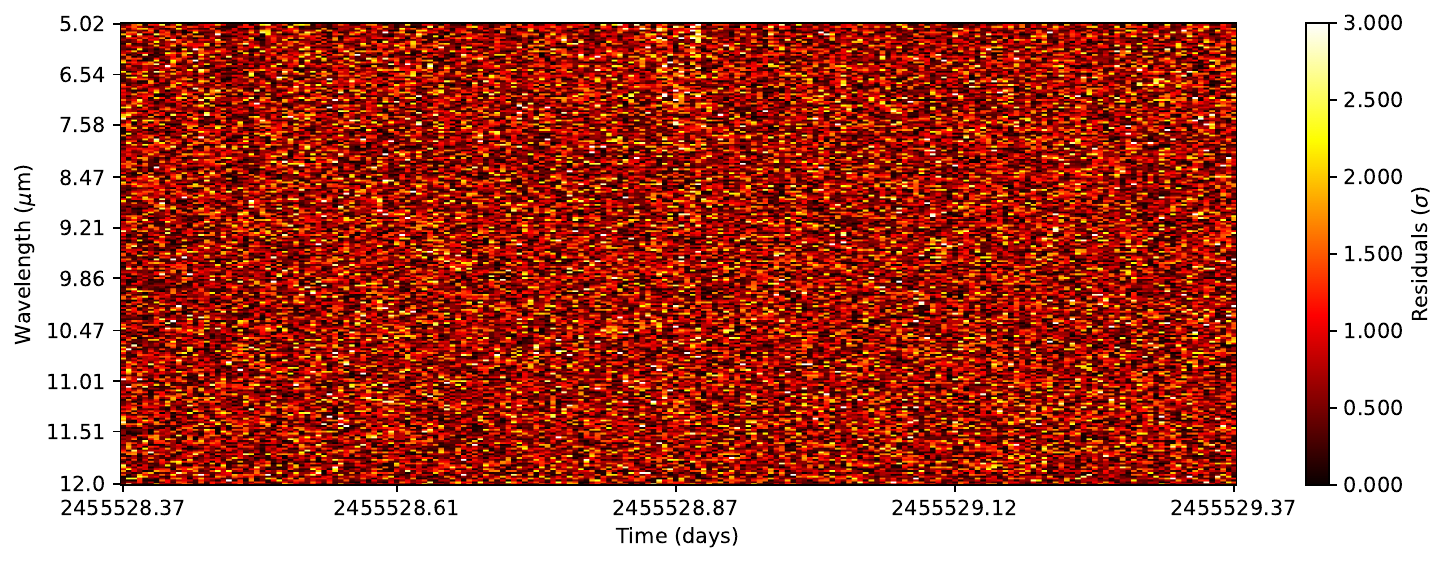}
    \caption{Simulated panchromatic light-curve observation of a WASP-43\,b exoplanet with JWST-MIRI (top) for our Scenario 2. Corresponding best-fit solution from the \textsc{ExPLOR} atmospheric retrieval (middle). Residuals between the simulated observations and the best fit solution and normalized by the observational uncertainties $\sigma$ (bottom).}
    \label{fig:wasp43_map_ggchem}
\end{figure*}

\begin{figure}[H]
\centering
    \includegraphics[width = 0.99\textwidth]{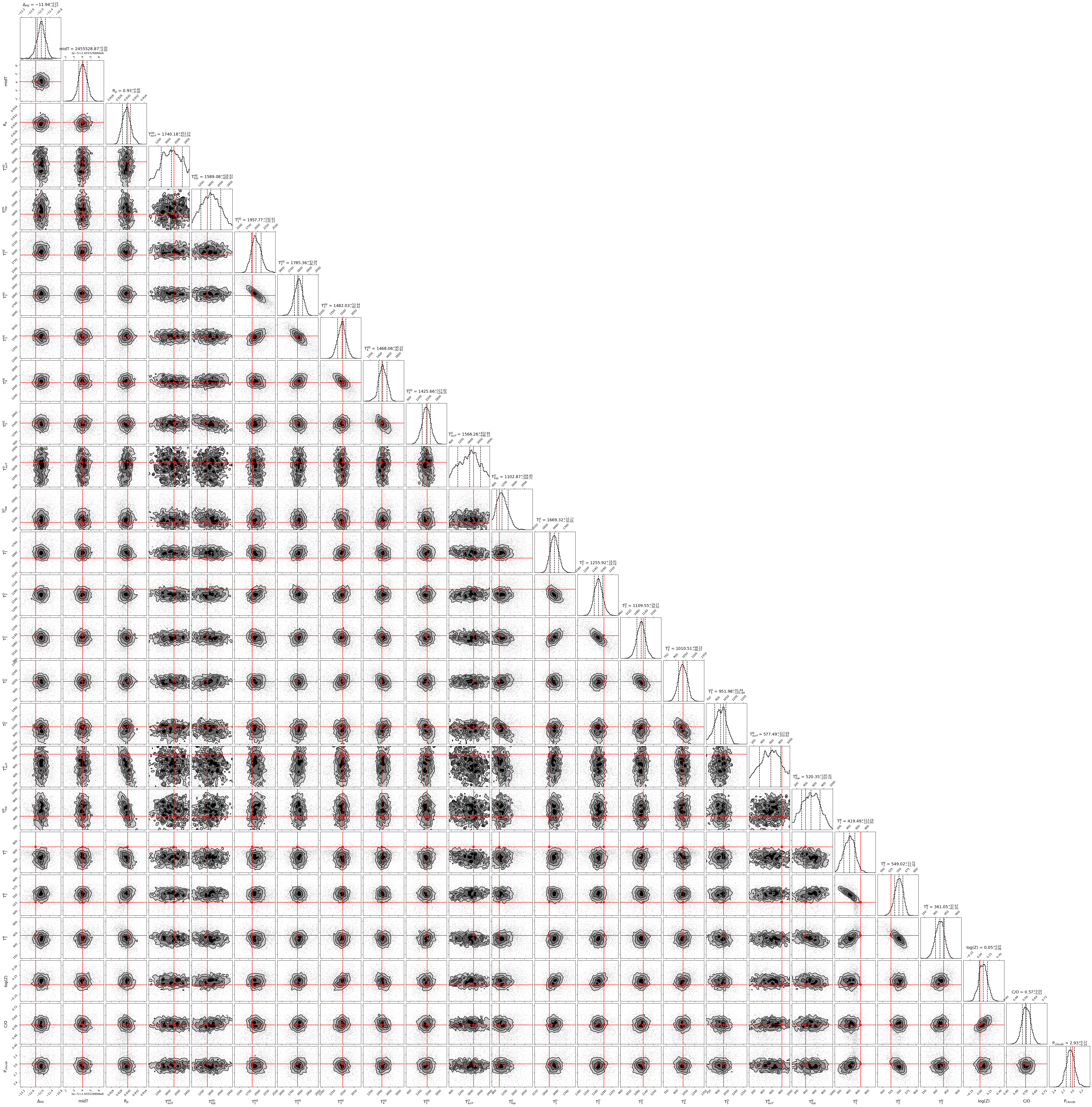}
    \caption{Full corner plot of the \textsc{ExPLOR} panchromatic light-curve retrieval on the WASP-43\,b like case for Scenario 2. The true value of this simulation is indicated by the red crosses. Note that the temperature parameters do not have ``true'' values per say as the location of the nodes is different in the forward and retrieval models. We therefore indicate the absolute temperature of the forward model profiles at the pressures set for the retrieved nodes.}
    \label{fig:wasp43_corner_JWST_ggchem}
\end{figure}

This appendix contains the complementary figures to the main article, Figure \ref{fig:wasp43_tp} to Figure \ref{fig:wasp43_corner_JWST_ggchem}. Figure \ref{fig:wasp43_tp} shows the thermal profiles obtained by the panchromatic light-curve retrievals for our simulations of a WASP-43\,b like atmosphere. Figure \ref{fig:wasp43_corner_JWST} shows the posterior distribution for the free chemistry retrievals (Scenario 1) obtained using \textsc{ExPLOR}. Figure \ref{fig:wasp43_map_ggchem} shows the simulated spectro-temporal flux maps from the simulation, retrieval and residuals in the chemical equilibrium case (Scenario 2). Figure \ref{fig:wasp43_corner_JWST_ggchem} shows the posterior distribution for Scenario 2. 

\vfill
\vfill
\clearpage

\clearpage

\twocolumngrid

\section{Appendix B: Setup of the Eclipse Light-curve retrievals}

\renewcommand{\floatpagefraction}{.99}%

\setcounter{figure}{0}
\renewcommand{\thefigure}{B\arabic{figure}}
\renewcommand{\theHfigure}{B\arabic{figure}}

To compare panchromatic light-curve retrievals to standard spectral retrievals, we produce a JWST-MIRI eclipse simulations of a WASP-43\,b-like planet using the hot-spot $T-p$ profile of Scenario 1 (see main text). More specifically, we compare the inversion strategy of \textsc{ExPLOR} to the more commonly employed two-step procedure involving light-curve reduction with a simplified transit model followed by a 1D spectral retrieval strategy (e.g., with \textsc{TauREx}). Here, the forward spectro-temporal map is produced using \textsc{ExPLOR}, coupling the three regions to ensure a 1D eclipse (i.e, there are no spatial variations in this example). We describe below the two retrieval strategies: \\

\begin{figure*}
\centering
    \includegraphics[width = 0.32\textwidth]{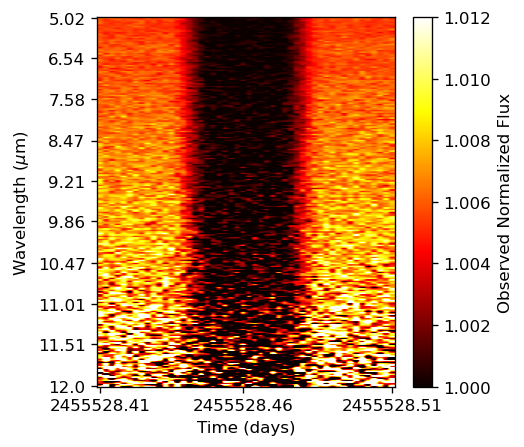}
    \includegraphics[width = 0.32\textwidth]{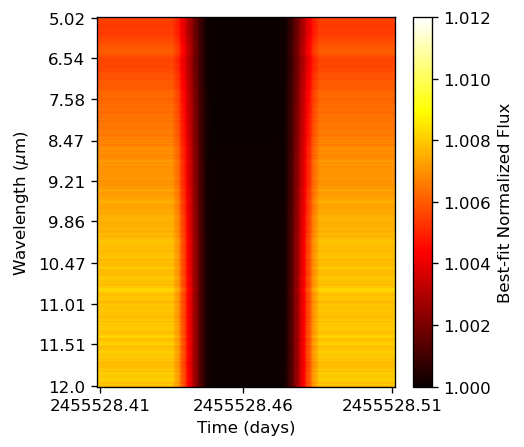}
    \includegraphics[width = 0.32\textwidth]{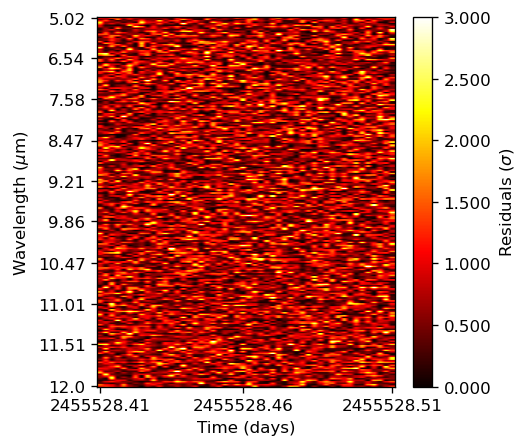}
    \includegraphics[width = 0.32\textwidth]{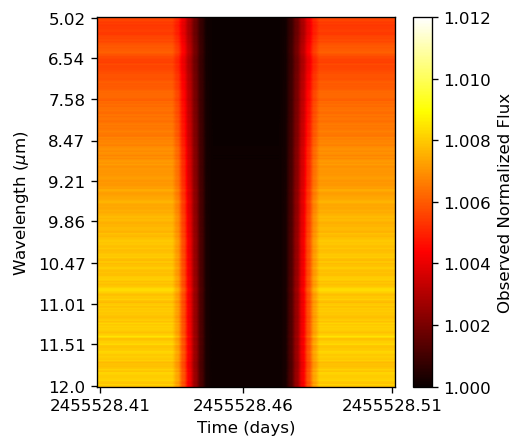}
    \includegraphics[width = 0.32\textwidth]{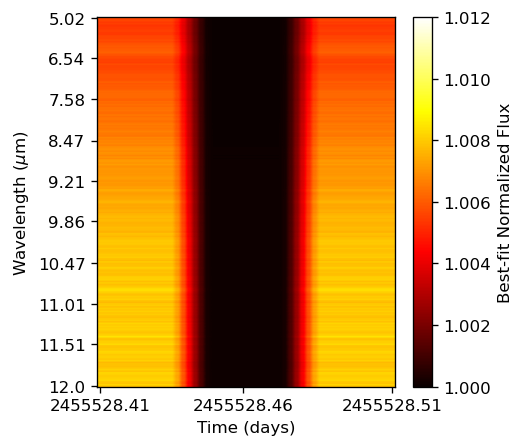}
    \includegraphics[width = 0.32\textwidth]{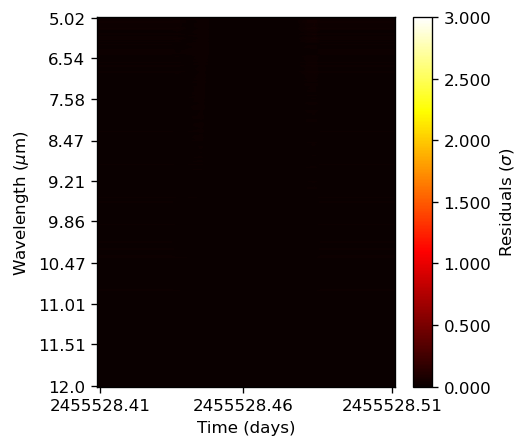}
    \caption{Panchromatic light-curve retrievals (ExPLOR) for an eclipse of a WASP-43\,b like planet when the data is scattered (top row) versus un-scattered (bottom row). A simulated spectro-temporal flux map convolved with a JWST-MIRI instrument model (left) serves as input to the retrievals. The retrieved best-fit maps obtained with \textsc{ExPLOR} (middle) and the residual maps (right) show the ability of the model to interpret such observations.}
    \label{fig:test_pure}
\end{figure*}

\underline{\it Panchromatic light-curve retrieval case} (\textsc{ExPLOR} retrieval): We recover a single thermal profile for all three regions (nightside, dayside, and hotspot) as well as the global abundances of H$_2$O and CO$_2$. We also fit for the planetary radius and mid-transit time, making a total of nine free parameters. We study noise propagation using two cases: a perfect case where the input spectro-temporal flux map {\it is not} scattered according to the uncertainties; and a realistic scenario where we repeat our retrievals for five normally scattered instances of the input spectro-temporal flux maps. As done previously, the inputs fed to \textsc{ExPLOR} are at native instrument resolution. We illustrate our panchromatic light-curve approach for the eclipse case in Figure \ref{fig:test_pure}, showing a scattered instance in the top row and the un-scattered case in the bottom row. \\

\underline{\it Standard reduction + retrieval case} (\textsc{TauREx} retrieval): To illustrate the standard practice of the literature, we start from the same spectro-temporal maps (left column of Figure \ref{fig:test_pure}), but now perform a light-curve reduction step to produce an eclipse spectrum. The spectrum is then fed to a 1D spectral retrieval. We first bin the simulated observation (i.e., the spectro-temporal maps) into 60 linearly separated wavelength channels -- this is typically done to increase the signal-to-noise in each light-curve. We then employ the \textsc{PoP} \citep{Changeat_2023} tool to fit each light-curve independently. In this fit, the orbital parameters are kept fixed (i.e., we assume those parameters can be obtained with a prior fit of the white-light curve, or via external measurements), and we model the astrophysical signal with a single free parameter: the eclipse depth. We add an instrument systematics model, composed of a single vertical offset free parameter. This offset parameter illustrates the addition of an independent wavelength-by-wavelength component, as would be the case in a real data reduction scenario \citep{tsiaras_hd209}. However, we here note that the addition of this parameter does not change our results, probably because the \textsc{ExPLOR} model did not include any instrument systematics commonly found in real data (i.e., short exponential ramps and long-term quadratic ramps). The inclusion of such effects can be explored in future works. Once the light-curves are fitted, we construct the eclipse spectrum by taking the median and variance of the day-side flux, as done by many reduction pipelines \citep{tsiaras_hd209, Bell_2022_eureka}. The information from this reduced spectrum is then inverted with the \textsc{TauREx3} retrieval framework using the base 1D eclipse model \citep{2019_al-refaie_taurex3, al-refaie_2021_taurex3.1}. As with the \textsc{ExPLOR} retrievals, we retrieve a 1D thermal profile using the same pressure nodes (i.e., with five free parameters), the abundance of H$_2$O and CO$_2$, and the planetary radius. \\

\begin{figure*}
\centering
    \includegraphics[width = 0.36\textwidth]{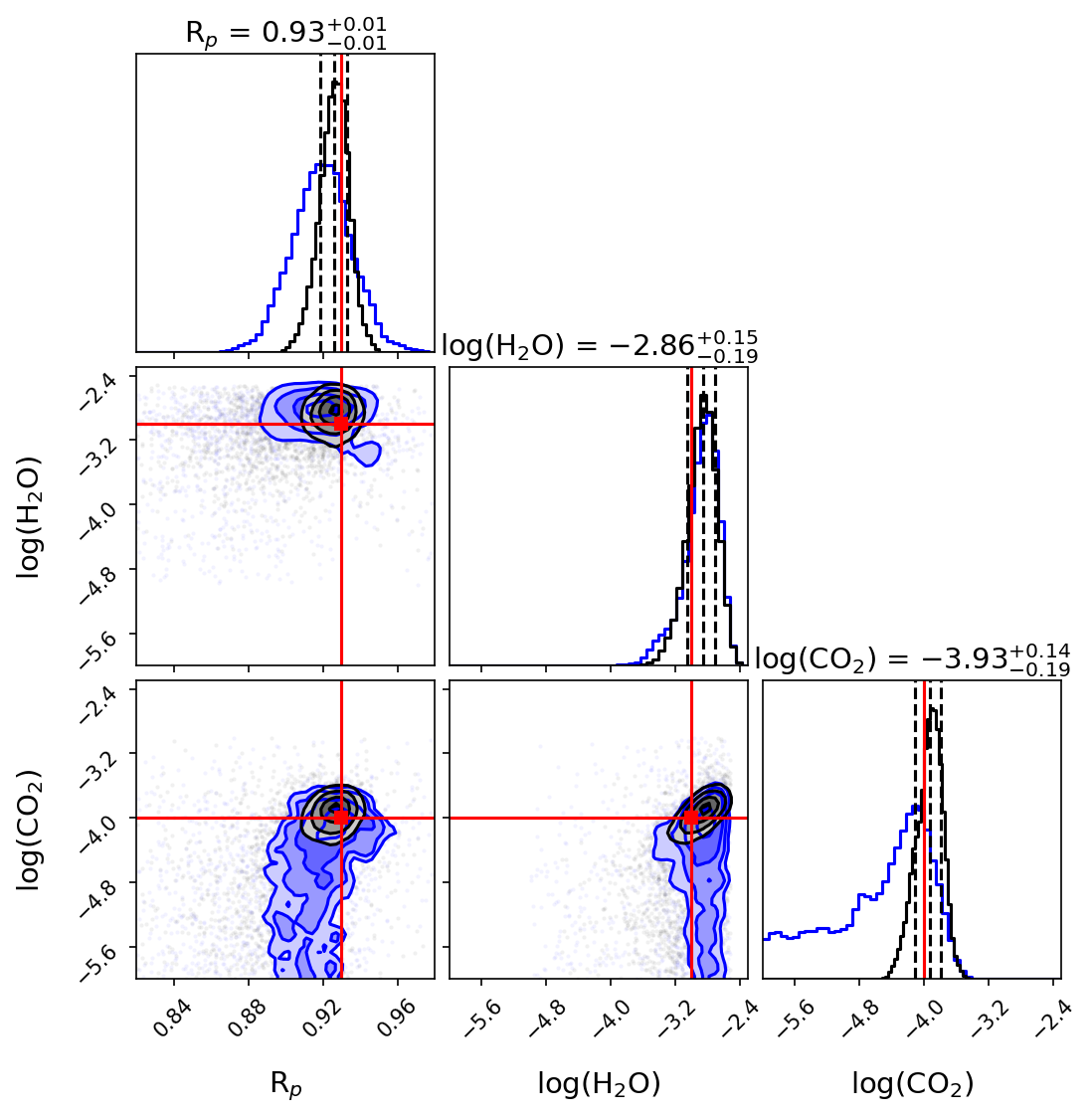}
    \includegraphics[width = 0.36\textwidth]{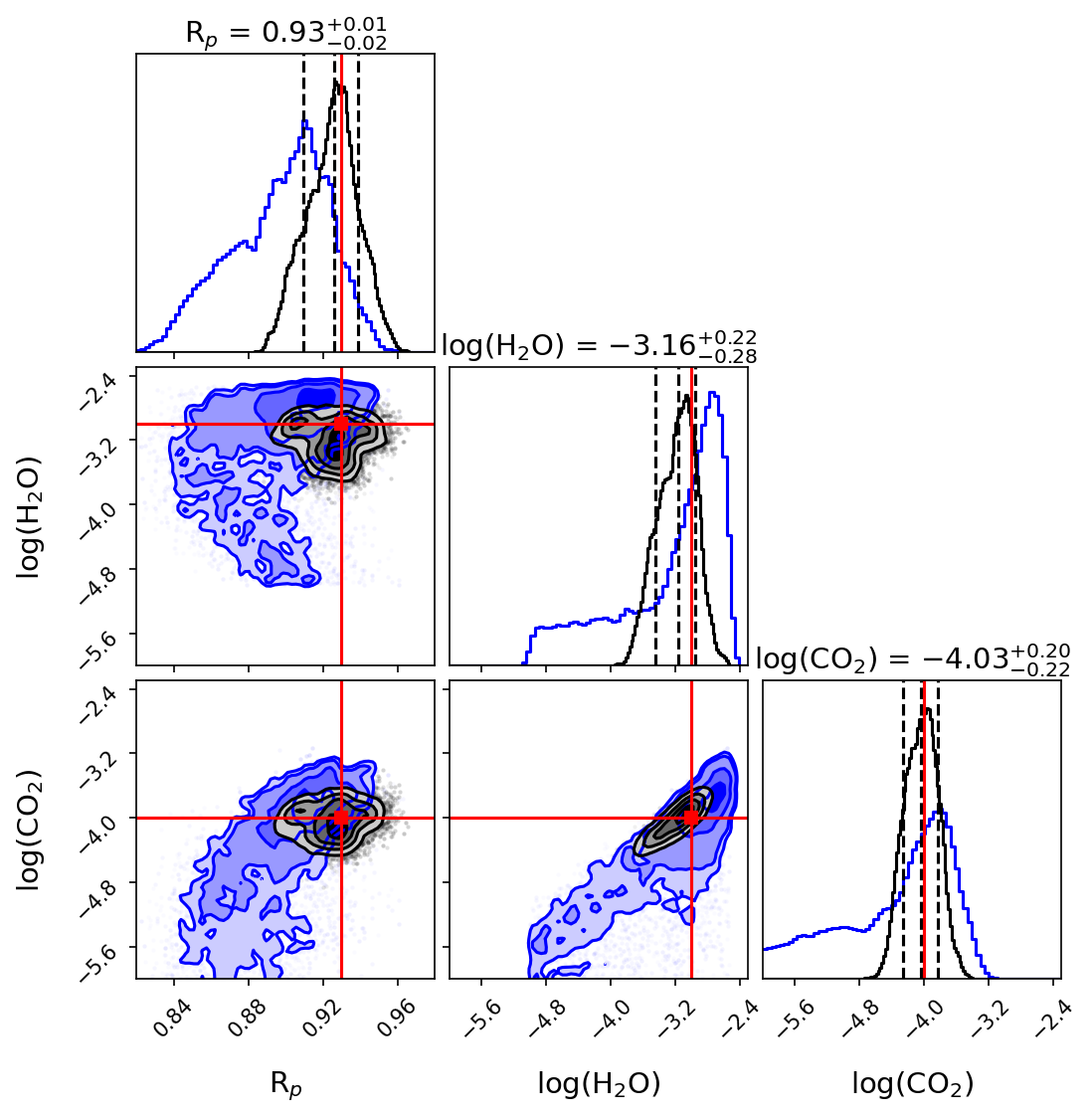}
    \includegraphics[width = 0.26\textwidth]{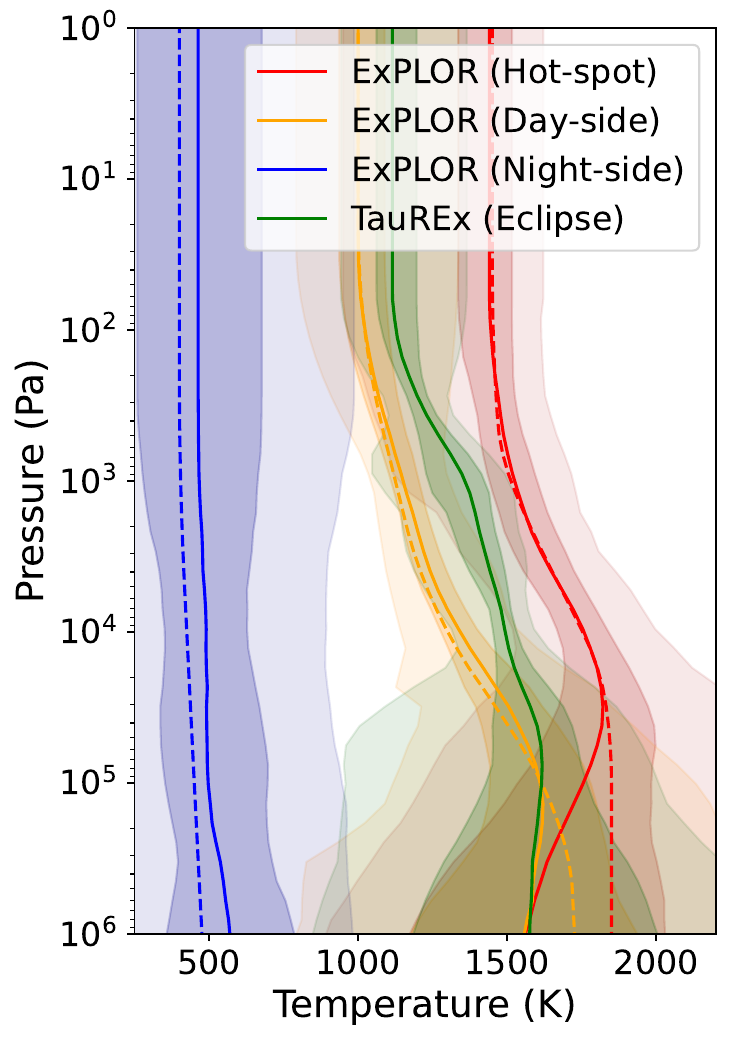}
    \caption{Corner plots (left: un-scattered input; middle: combination of five independently scattered inputs) from the \textsc{ExPLOR} (black) and the \textsc{TauREx} (blue) retrievals obtained for the 1.5D (i.e, spatially varying atmosphere) JWST eclipse examples, and retrieved $T-p$ structure (right) for the un-scattered case. Similar results to the 1D case of Figure \ref{fig:wasp43_spectra_corners} are found but the \textsc{ExPLOR} retrieval allows to perform eclipse mapping while the \textsc{TauREx} retrieval is 1D. The retrieval \textsc{TauREx} $T-p$ profile can only obtains an average of the dayside and hotspot profiles, but it is reassuring to find that no significant bias is introduced in this case. By design, the \textsc{ExPLOR} retrieval accounts for 3D biases in eclipse and transit \cite[as described in e.g., ][]{Feng_2016_inomogeneou, Caldas_2019, Taylor_2020, MacDonald_2022}), extracting more realistic information from the data than transitional 1D retrievals.}
    \label{fig:test_pure_temperature}
\end{figure*}

In addition to those tests, we performed eclipse retrievals on the data from Scenario 1 where spacial inhomogeneities are present (see main text). Since the $T-p$ profile is different for each region, traditional 1D spectral retrievals should struggle more (a phase-dependent behaviour is visible even when only considering the eclipse only). To partially account for this -- as would be done with real data -- the light-curve stage is performed using a sine phase-curve model composed of three free parameters: the night-side flux, the day-side flux, and the hot-spot phase offset. The \textsc{ExPLOR} retrieval is performed using an independent $T-p$ profiles for each regions. The rest proceed as described with the previous eclipse tests. Figure \ref{fig:test_pure_temperature} shows the posterior distributions and retrieved profiles for this case, highlighting that 1) chemical parameters of the 1D retrieval are not biased in this case, 2) information dilution also occurs (similar conclusion to the 1D case), and 3) the $T-p$ profile in the eclipse retrieval converges towards a mean value between the dayside and hotspot profiles.

\end{document}